\documentclass[preprint,3p,11pt]{elsarticle}

\usepackage{amssymb}
\usepackage{amsmath}
\usepackage[colorlinks,citecolor=blue,linkcolor=blue,urlcolor=blue,bookmarksnumbered=true,pdfauthor=author]{hyperref}
\usepackage{cleveref}
\usepackage{float}
\usepackage{graphicx}
\usepackage{bm}
\usepackage{booktabs}
\usepackage{graphicx}%
\usepackage{multirow}%
\usepackage{amsmath,amssymb,amsfonts}%
\usepackage{amsthm}%
\usepackage{mathrsfs}%
\usepackage[title]{appendix}%
\usepackage{xcolor}%
\usepackage{textcomp}%
\usepackage{manyfoot}%
\usepackage{booktabs}%
\usepackage{algorithm}%
\usepackage{algorithmicx}%
\usepackage{algpseudocode}%
\usepackage{listings}%
\usepackage{hyperref}
\usepackage{float}
\usepackage{graphicx}
\usepackage{bm}
\usepackage{booktabs}
\usepackage{natbib}
\usepackage{caption}
\newcounter{appendixfigure}
\renewcommand{\theappendixfigure}{S\arabic{appendixfigure}}
\usepackage{lineno}

\usepackage{fancyhdr}
\fancypagestyle{mypage}{
    \fancyhf{} 
    \fancyhead[R]{\thepage} 
}
\fancypagestyle{plain}{
    \fancyhf{}
    \fancyfoot[C]{\thepage} 
}
\usepackage{lineno}

\journal{Elsevier}
\usepackage{times}
\begin{document}
\begin{frontmatter}



\title{\textbf{Imaging neutron radiation-induced defects in single-crystal chemical vapor deposition diamond at the atomic level}} 


\author[label1,label2,label6]{Jialiang Zhang} 
\author[label1,label3,label6]{Futao Huang} 
\author[label1,label2]{Shuo Li} 
\author[label1,label2]{Guojun Yu}
\author[label1,label2]{Zifeng Xu}
\author[label4]{Lifu Hei}
\author[label4]{Fanxiu Lv}
\author[label5]{Aidan Horne}
\author[label1,label3,label5]{Peng Wang\corref{cor1}}
\author[label1,label2]{Ming Qi\corref{cor1}}

\affiliation[label1]{organization={National Laboratory of Solid State Microstructures},
            addressline={Nanjing University}, 
            city={Nanjing},
            postcode={210093}, 
            country={China}}
\affiliation[label2]{organization={School of Physics},
            addressline={Nanjing University}, 
            city={Nanjing},
            postcode={210093}, 
            country={China}}
\affiliation[label3]{organization={College of Engineering and Applied Sciences and Collaborative Innovation Center of Advanced Microstructures},
            addressline={Nanjing University}, 
            city={Nanjing},
            postcode={210023}, 
            country={China}}
\affiliation[label4]{organization={School of Materials Science and Engineering},
            addressline={University of Science and Technology Beijing}, 
            city={Beijing},
            postcode={100083}, 
            country={China}}
\affiliation[label5]{organization={Department of Physics},
            addressline={University of Warwick}, 
            city={Coventry},
            postcode={CV4 7AL}, 
            country={United Kingdom}}
\fntext[label6]{These authors contributed equally}          

\cortext[cor1]{Corresponding author.}

\begin{abstract}
Diamond’s exceptional properties make it highly suited for applications in challenging radiation environments. Understanding radiation-induced damage in diamond is crucial for enabling its practical applications and advancing materials science. However, direct imaging of radiation-induced crystal defects at the atomic to nanometer scale remains rare due to diamond's compact lattice structure. Here, we report the atomic-level characterization of crystal defects induced by high-flux fast neutron radiation (up to $3\times10^{17}$n/$\mathrm{cm}^2$) in single-crystal chemical vapor deposition diamonds. Through Raman spectroscopy, the phase transition from carbon $\mathrm{sp}^3$ to $\mathrm{sp}^2$ hybridization was identified, primarily associated with the formation of dumbbell-shaped interstitial defects, which represent the most prominent radiation-induced defects. Using electron energy loss spectroscopy and aberration-corrected transmission electron microscopy, we observed a clustering trend in defect distribution, where $\mathrm{sp}^2$-rich clusters manifested as dislocation cluster structures with a density up to $10^{14}$ $\mathrm{cm}^{-2}$. Lomer-Cottrell junctions with a Burgers vector of 1/6$\langle 110 \rangle$ were identified, offering a possible explanation for defect cluster formation. Radiation-induced point defects were found to be dispersed throughout the diamond lattice, highlighting the widespread nature of primary defect formation. Vacancy defects, along with $\langle 111 \rangle$ and $\langle 100 \rangle$ oriented dumbbell-shaped interstitial defects induced by high-dose neutron irradiation, were directly imaged, providing microscopic structural evidence that complements spectroscopic studies of point defects. Dynamical simulations combined with an adiabatic recombination-based crystal damage model, provided insights into the correlation between irradiation dose and resulting crystal damage. These findings advance our understanding of neutron-induced radiation damage mechanisms in diamond and contribute to the development of radiation-resistant diamond materials.

\end{abstract}




\end{frontmatter}



\newpage
\restoregeometry

   \tableofcontents
\section{Introduction}

Chemical vapor deposition (CVD) diamond has emerged as a potentially compelling candidate material for precise semi-conductor devices and sensitive detectors operating within severe irradiated environments, assisting humanity in exploring uncharted domains in high energy physics, fusion reactors, and space explorations\cite{Field2012}\cite{Childress2006}\cite{Tapper2000}\cite{Narayan2016}\cite{Pace2000}\cite{Ogawa2023}. Benefiting from a wide band gap of 5.5 eV\cite{Wort2008} and high displacement energy of approximately 43 eV\cite{Koike1992}, diamond is endowed with incomparable resistance to radiation.

Understanding the radiation-induced damage in the atomic structure of materials is important for improving device design and discovering new materials with enhanced properties\cite{ZHANG2017221}\cite{Zhang2022}. Also, valuable insights can be provided to analyze the interaction mechanism between radiation and materials.
Concerning the radiation effect on diamond, the majority of research is intertwined with characterizing device performance\cite{Zou2020}\cite{DeBoer2007}\cite{Liu2020}, the current understanding of lattice damage caused by irradiation in diamond is still limited\cite{Nordlund2018}.

Direct atomic-scale characterization of lattice damage in diamond remains unreported, particularly at high-dose irradiation levels, due to the challenges posed by the dense packing of carbon atoms in the diamond lattice and the rarity of high-dose irradiation experiments. Transmission electron microscopy (TEM), equipped with advanced aberration correction and scanning techniques, provides a powerful tool for overcoming the challenges posed by the high atomic density in the diamond lattice. With sub-angstrom atomic resolution, TEM enables clear imaging of microscopic structures, such as dumbbells within the \{110\} plane of natural diamond. E. Olivier et al. used aberration-corrected TEM to investigate the atomic structure and local chemistry of platelet defects in diamond\cite{Olivier2018}, while P. Nemeth et al. employed scanning transmission electron microscopy (STEM) to elucidate the complex nanostructure of diamond\cite{Nemeth2020}\cite{Garvie2014}, E.K. Nshingabigwi studied ion implantation in diamond with TEM\cite{nshingabigwi_electron_2014}. These advancements have laid a crucial foundation for atomic-level characterization of diamond structures, facilitating similar studies on radiation-induced damage.

To address the challenge posed by the rarity of high-dose irradiation experiments, single-crystal CVD (sc-CVD) diamond was subjected to high-dose fast neutron irradiation to investigate irradiation-induced damage in diamond crystals. Initially, several sc-CVD diamond wafers, each measuring $6 \times 6 \times 0.3$ $\mathrm{mm}^3$, were synthesized in the laboratory with the goal of developing a new type of high-energy particle detector. These samples were then exposed to fast neutrons at the IBR-2 facility in Dubna, Russia\cite{Bulavin2018}. The total fast neutron fluence reached $3.3 \times 10^{17}$ n/cm², surpassing the radiation dose expected for diamond materials used in next-generation high energy particle colliders\cite{Rutherfoord2012}\cite{8116686}\cite{Bejar2020}. Due to the significant non-ionizing energy loss cross-section of fast neutrons\cite{DeBoer2007}, the high-dose neutron irradiation experiment provided ideal conditions for investigating irradiation-induced damage in diamond.

In this work, Raman spectroscopy was employed to identify the intrinsic vibrational modes in diamond induced by irradiation. Electron energy loss spectroscopy (EELS) was utilized to explore potential carbon phase transitions and changes in local chemistry within the diamond material. Subsequently, aberration-corrected scanning transmission electron microscopy (STEM) with a high-angle annular dark field (HAADF) detector, along with high-resolution transmission electron microscopy (HRTEM) in the conventional model, were used to directly image neutron-induced defects and structural changes at the atomic scale. The morphological characteristics and distribution of these defects were examined, providing valuable insights into the irradiation-induced damage and its underlying mechanisms in diamond.

\section{Experimental}
\subsection{Synthetic of sc-CVD diamond and neutron irradiation experiment}
Freestanding single-crystal diamond samples were fabricated on a commercial High Pressure High Temperature (HPHT) synthetic single-crystal diamond seed oriented along the [001] direction. A high-power direct current arc plasma Jet chemical vapor deposition technique\cite{Hei2012} was employed to prepare large-sized high-quality sc-CVD diamond in this research, which offers the advantage of reduced cost. To minimize nitrogen incorporation, we implemented an optimized growth process, which included the use of a high-sealing system in the CVD growth chamber to reduce contamination, high-purity source gases to limit nitrogen levels from the gas phase, and a smaller (10 kW) direct-current (dc) arcjet system operating in blow-down mode (without gas recycling) to further enhance the purity of the diamond. Through these optimizations, the nitrogen content in the single-crystal CVD diamond was controlled to below 500 ppb. Subsequently, when the CVD layer had grown to a thickness of 1 millimeter, laser cutting was employed to separate the freestanding sc-CVD diamond from the seed, followed by a polishing process. The samples measured $6\times6\times$ $\mathrm{mm}^2$in surface area, with thicknesses ranging from 270 $\mathrm{\mu}m$ to 350 $\mathrm{\mu}m$. The surface normal was oriented along the [001] crystallographic direction. Then Ti-W-Au electrodes were coated on the both upper and lower surfaces of these sc-CVD diamond samples, the electrodes were connected to the electronics and voltage sources by wire bonding, transforming these samples into sc-CVD diamond electronic devices under test (DUT). Six sc-CVD DUTs were placed on a neutron beamline in the IBR-2M reactor in Dubna, specifically designed for irradiation experiments\cite{bulavin_irradiation_2015}. These DUTs were exposed to a fast neutron flux over a 12-day period, reaching a total fluence of $3.3 \times 10^{17} \mathrm{n/cm^2}$. This fluence corresponds to the irradiation level expected for diamond detectors over ten years of operation in next-generation high-luminosity collider physics experiments\cite{Rutherfoord2012}. During irradiation experiments, the sample temperature was actively monitored and maintained below 50°C, typically slightly above ambient temperature, through the extraction ventilation system\cite{bulavin_irradiation_2015}. The readout signals of the single-crystal CVD DUTs were continuously monitored throughout the irradiation experiment, and the DUTs remained fully operational under high-dose irradiation. The irradiated diamond devices were retrieved and subjected to an acid bath to remove the surface metal electrodes after the irradiation experiment. Identical coating and cleaning methods were applied to the non-irradiated single-crystal CVD diamond samples in our laboratory. Both sets of samples subsequently underwent identical characterization experiments to facilitate direct comparison.
\subsection{Transmission electron microscopy and electron energy loss spectroscopy}
Following the irradiation experiment, transmission electron microscopy was employed to elucidate the lattice structure of the irradiation-induced damage within the diamond.
\subsubsection{Focused ion beam}
To facilitate TEM sample preparation of single-crystal diamond, a focused ion beam (FIB) in situ lift-out and milling method was employed. The method we used, as described by D. P. Hickey \cite{hickey_cross-sectional_2006} and E. J. Olivier\cite{Olivier2018}, had been successful applied in TEM observation of damage and defects in diamond atomic structures. Ga+ ions with an energy of 30 keV and a current of 1000 pA were used to mill two wedges, leaving a 10 $\mathrm{\mu m}$ long, 5 $\mathrm{\mu m}$ deep, and 2 $\mathrm{\mu m}$ thick diamond slice in between. The sample was then lifted off from the top and thinned using Ga ions at a grazing angle (almost parallel to the sample surface) to minimize energy transfer. In the first stage, the sample was thinned to a thickness of less than 1 $\mathrm{\mu m}$\. It was then further thinned to about 500 nm by reducing the Ga+ ion beam energy and using a lower current of 100 pA. The thickness was subsequently decreased to around 100 nm with a further reduced ion beam voltage of 2 keV while maintaining the current at 100 pA. The use of 2 keV Ga ions can reduced the possible FIB-induced damage layer to less than 2 nm \cite{hickey_cross-sectional_2006}. A final thinning and cleaning process was performed using Ga+ ions at 500 V and 50 pA, yielding a sample thickness of around 50 nm and ensuring that Ga-induced damage did not affect our observations. These steps effectively minimized the amorphization effect caused by the FIB process to just a few atomic layers or less, allowing for accurate characterization of the irradiation-induced defects\cite{rubanov_study_2011}\cite{fib2}.To rule out the influence of the sample preparation process, we prepared unirradiated diamond samples using the same method. EELS and TEM characterization revealed no amorphous or sp² peaks, and the lattice structure remained intact, indicating that the preparation process introduced minimal damage. Although we cannot completely rule out the possibility of minor structural changes induced by Ga ions, such damage was not detected by electron energy loss spectroscopy (EELS). Furthermore, by comparing irradiated and unirradiated samples prepared under identical conditions, we attribute any identical TEM or spectroscopic results to the sample preparation process. These artifacts are excluded from our analysis, and we focus solely on the differences before and after irradiation, thereby ensuring the reliability of our results.
\subsubsection{Transmission electron microscopy}
TEM experiments and instruments are illustrated in this section. Atomic-resolution STEM-HAADF images of irradiated diamond were obtained on a double spherical aberration-corrected STEM/TEM FEI Titan G2 60-300 instrument at 300kV accelerating voltage with a field emission gun. The probe convergence angle on the Titan electron microscope was 22.5 mrad, and the angular range of the HAADF detector was from 79.5 to 200 mrad. Also,
conventional HRTEM images were recorded with an FEI Tecnai TF20 field emission gun TEM in high-resolution mode. The accelerating voltage was set at 200kV. 
TEM images were processed and enhanced using Fast Fourier Transform (FFT) filtering with ImageJ software\cite{schneider_nih_2012} to reduce noise and improve clarity.
For lattice structure and dislocation analysis, we utilized Geometric Phase Analysis (GPA) measurements based on the open-source program Strainpp. The implementation of the GPA algorithm in Strainpp follows the methodology detailed by Martin Hÿtch\cite{Hytch1998}.

Electron energy-loss spectra (EELS) records of both neutron-irradiated diamond and non-irradiated diamond were on a Gatan 966 EEL spectrometer, the collection aperture was set to 2.5 mm and the camera length was 29.5 mm.

\subsection{Simulations}
\subsubsection{Simulations of damage process}
The primary and secondary damage caused by neutron irradiation within diamond was simulated using Monte Carlo (MC) and Molecular Dynamics (MD) methods, respectively, enhancing our understanding of neutron irradiation processes. The transport and damage behavior of fast neutrons within diamond were modeled using GEANT4\cite{Agostinelli2003}, employing the QGSP-BERT-HP physics list for its precise description of neutron-carbon scattering at low and intermediate energies.

In the case of recoil diamond atoms with energies in the hundreds of keV range, interactions were described by hard sphere scattering with the Born-Mayer potential\cite{Born1932} for energies below $10^3$ eV, and inverse square approximation using a screened Coulomb potential fit for energies between $10^3$ and $10^5$ eV\cite{was2016fundamentals}. Building upon this principle, the Large-scale Atomic/Molecular Massively Parallel Simulator (LAMMPS)\cite{LAMMPS} molecular dynamics program was employed to investigate the dynamic interactions between energetic primary knock-on atoms (PKA) and diamond crystals.
A diamond lattice comprising 1,000,000 atoms in a $50 \times 50 \times 50$ unit cell configuration was constructed. The Tersoff potential\cite{Tersoff1988} was used to model the atomic interactions between carbon atoms, supplemented by the ZBL potential\cite{Ziegler1985} for short-range interactions. Periodic boundary conditions were applied. The system was initially relaxed for 10 ps in the NVT ensemble, followed by irradiation in the NVE ensemble. The MD time step was set to less than 0.005 $a_0$.

\subsubsection{Simulations of TEM characterizations}
TEM simulations of lattice defects derived from molecular dynamics simulations enable direct comparison with defects observed in irradiated diamond samples, elucidates the dynamic mechanisms underlying the formation of specific defects.
The STEM-HAADF simulations were done using the abTEM\cite{madsen_abtem_2021} software package. The accelerating voltage for the simulations along the $\langle 100 \rangle$ zone axes was 300 keV. The imaging parameters were fixed for all simulations, with a semiconvergence angle of 22.5 mrad, an inner collection angle of 79.5 mrad, and an outer collection angle of 200 mrad. The sampling for the scan was set just below the Nyquist frequency for the optical setup, and then interpolation was done to remove pixelation from the image.
The HRTEM simulation was performed using the open-source software clTEM\cite{PETERS2021113364} under conditions consistent with the experiment, with an accelerating voltage of 200 kV. 
The Cs value was set to 1.2 mm, aligning with the experimental conditions employed using the TF20 field emission gun TEM. The input lattice file used was generated by molecular dynamics  simulation, which includes atomic vacancies in the form of dumbbell defects caused by neutron irradiation.

\section{Results and Discussion}

\subsection{Non-intrinsic vibrational modes by Raman spectroscopy}
As a result of irradiation effects, defects introduced by incident particles lead to altered atomic arrangements in single-crystal CVD diamond, forming localized new crystal structures. With these structures, new lattice vibrational modes emerge within the diamond crystal. Using Raman spectroscopy, we characterized these non-intrinsic vibrational modes to investigate the manifestations of irradiation-induced defects in diamond. Raman spectroscopy measurements were performed using a T64000 spectrometer equipped with a 532 nm laser. The laser power on the sample was maintained at approximately 100 mW, and the spectra were collected over a wavenumber range of 1000 $\mathrm{cm}^{-1}$ to 1800 $\mathrm{cm}^{-1}$ with a step size of 0.5 $\mathrm{cm}^{-1}$. The Raman spectra of irradiated single-crystal diamond, compared to the unirradiated sample, can be seen in Fig. 1(a). Before irradiation, the diamond sample exhibited a single prominent Raman peak at 1332.5 $\mathrm{cm}^{-1}$, with a full width at half maximum (FWHM) of approximately 2 $\mathrm{cm}^{-1}$. After irradiation, the diamond peak remains visible but exhibits a reduced intensity, a slight shift to 1331.5 $\mathrm{cm}^{-1}$, and an increased FWHM of approximately 4.3 $\mathrm{cm}^{-1}$. This suggests that, even under high-dose neutron irradiation, single-crystal CVD diamond retains substantial long-range order.The Raman spectrum showed a broad structural band in the 1100-1200 $\mathrm{cm}^{-1}$ range. As discussed in studies\cite{poklonski_magnetic_2023} and our earlier work\cite{Liu2020}, this band likely results from the superposition of characteristic peaks at 990, 1008, 1120, and 1235 $\mathrm{cm}^{-1}$, which correspond to phonon frequencies at critical points in the Brillouin zone (LA(L), LA(K), TO(W), TO(K), LO(K), and LO(L)). These features are attributed to the phonon confinement effect caused by irradiation-induced intrinsic defects, which reduce the phonon mean free path and modify the Raman spectrum. The narrow peak at $\sim$ 1360 $\mathrm{cm}^{-1}$ is likely due to localized vibrational modes of radiation-induced point defects, such as vacancies, interstitials, or defect complexes. This interpretation is supported by previous studies on ion-implanted and fast neutron irradiated diamonds, where similar narrow peaks in the 1350-1600 $\mathrm{cm}^{-1}$ range have been observed and attributed to localized defect vibrations \cite{khomich_probing_2020}\cite{poklonskaya_raman_2015}\cite{poklonskaya_raman_2013}. Additionally, the intensity and position of these peaks are known to depend on annealing temperature.

An additional broad peak around 1460 $\mathrm{cm}^{-1}$ is likely a superposition of multiple vibrational modes, with the 1445 $\mathrm{cm}^{-1}$ peak possibly related to diamond-like carbon, which has been observed in ion implantation damage in diamond \cite{leech_effect_2004}. Calculations suggest that peaks at 1461 $\mathrm{cm}^{-1}$ and 1495 $\mathrm{cm}^{-1}$ may originate from local vibrational modes of di-interstitials\cite{goss_self-interstitial_2001}. Additionally, in ion-implanted diamonds, single vacancies appear in the Raman spectrum as a series of bands in the range of 1420–1500 $\mathrm{cm}^{-1}$\cite{poklonskaya_raman_2013}, vacancy-related peaks at 1420 $\mathrm{cm}^{-1}$ and 1490 $\mathrm{cm}^{-1}$\cite{kalish_nature_1999}. Therefore, the presence of this band can be attributed to radiation-induced local defects. A strong and narrow vibrational mode at 1631 $\mathrm{cm}^{-1}$ is generally attributed to split interstitial defects, where two carbon atoms occupy a single atomic position, potentially aligned along the ⟨100⟩ crystal direction. This is a characteristic feature of the Raman spectra of radiation-damaged diamond\cite{prawer_raman_2000}\cite{orwa_raman_2000}. However, different hypotheses have also been proposed. For example, some researchers suggest that this band may originate from $\mathrm{sp}^2$-hybridized carbon in various forms\cite{ferrari_interpretation_2000}. Experiments on irradiated ultrafine detonation nanodiamonds (DND) have shown that this band can decompose into multiple peaks upon annealing, suggesting an interpretation distinct from split interstitial defects, indicating its potential origin from bulk defects in nanodiamonds\cite{khomich_features_2019}.

\subsection{Concentration and distribution of neutron radiation-induced $\mathrm{sp}^2$ structure}
To investigate the concentration and distribution of small $\mathrm{sp}^2$ structure and interstitial defects with enhanced spatial resolution, focused ion beam (FIB) techniques were employed to prepare cross-sectional samples from diamond bulk for TEM characterization, enabling detailed examination of neutron radiation-induced defects across multiple dimensions.
On the FIB-prepared TEM samples, Electron Energy Loss Spectroscopy (EELS) was employed to investigate irradiation-induced carbon bonding transitions and their spatial distribution. Square regions with side lengths of 100 nanometers were selected on neutron-irradiated diamond, and equivalent-sized regions were also chosen on non-irradiated diamond for comparative analysis. Fig. 1(b) and Fig. 1(c) illustrate the high-energy and low-energy electron loss spectra observed in diamond, respectively.In neutron-irradiated diamond, a specific peak appeared around 285 eV, and a significant decrease in peak intensity was detected at around 33 eV and 292 eV. The peak at 285 eV corresponds to the transition of carbon $1s$ to the $\pi^*$ state, representing the $\mathrm{sp^2}$ bonding in C-C\cite{Muller1993}. The 292 eV signal in the core energy loss spectrum of diamond indicates the $\mathrm{sp^3}$ bonding in the C-C structure as the transition from the carbon $1s$ to the $\sigma^*$ state\cite{Muller1993}. Furthermore, the peak around 33 eV with a shoulder at 23 eV in the low-energy loss spectrum corresponds to the plasma peak from $\omega_{2D}$ $\mathrm{sp^3}$ bonding in the crystalline diamond\cite{Rani2018}. 
These results demonstrate that neutron irradiation induces the occurrence of $\mathrm{sp^2}$ bonding in diamond, whereas non-irradiated diamond shows a low presence of $\mathrm{sp^2}$ amorphous or graphite-like structure. the $\mathrm{sp^2}$ to $\mathrm{sp^3}$ comparison $\mathrm{I}(\pi^*)/\mathrm{I}(\sigma^*)$ hybridization ratio in imaging region can be deducted as 5\% based on Ratio height methods. To investigate the spatial distribution of the $\mathrm{sp^2}$ component, energy loss signals associated with $\mathrm{sp^2}$ bonding in the 283–287 eV range were extracted from the spectrum and spatially mapped based on signal intensity, revealing their locations within the diamond. As shown in the mapping image in Fig. 1(d), the spatial distribution of $\mathrm{sp^2}$ bonding in irradiated diamond is highly localized, with concentrations appearing within specific regions spanning approximately tens of nanometers. To further analyze these areas, we selected regions of high $\mathrm{sp^2}$ concentration (highlighted in red) alongside normal regions (highlighted in blue) for comparison of their energy loss spectra, presented in Fig. 1(e). The region with concentrated $\mathrm{sp^2}$ bonding exhibits a more pronounced peak at 285 eV and demonstrates an $\mathrm{sp^2}$ to $\mathrm{sp^3}$ ratio, $\mathrm{I}(\pi^*)/\mathrm{I}(\sigma^*)$, of roughly 16\%, while in the non-concentrated sp² region (blue square in Fig. 1d),  this ratio is around 3\%. The quantification of the sp²/sp³ ratio was performed using the two-window method\cite{bruley_quantitative_1995}, where the $\mathrm{sp^2}$ fraction $x$ is calculated as: $$\left(I_{\boldsymbol{\pi} *}^{u} / I_{\boldsymbol{\sigma} *}^{u}\right) /\left(I_{\boldsymbol{\pi} *}^{s} / I_{\boldsymbol{\sigma} *}^{s}\right)=3 x /(4-x)$$,where $u$ represents the unknown test sample, and $s$ denotes the standard sample.
\begin{figure}[H]
\label{fig1}
\centering
\includegraphics[width=\linewidth]{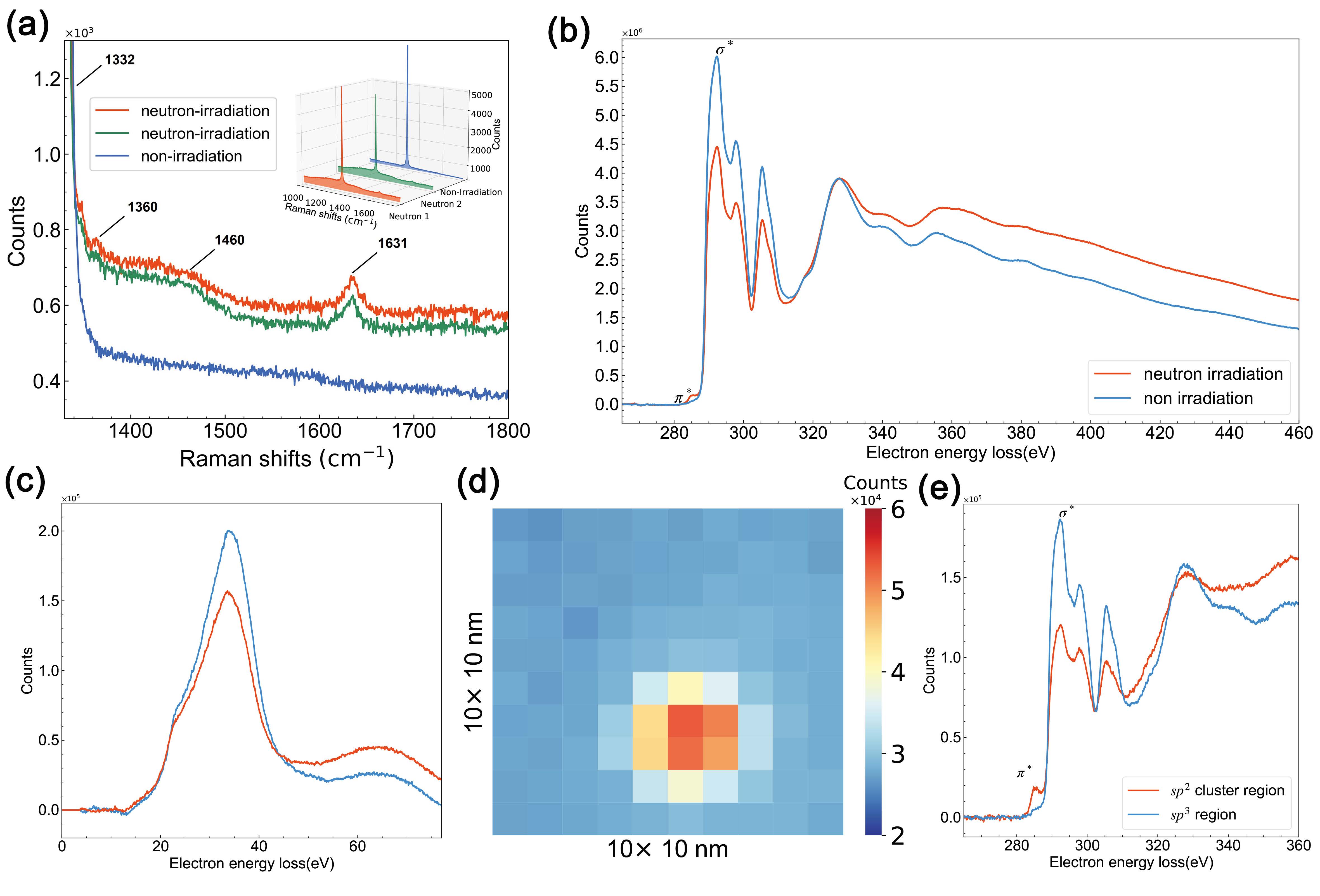}
\caption{Raman and electron energy loss spectroscopy of neutron-irradiated sc-CVD diamond\textbf{a,} Raman spectra of neutron-irradiated and non-irradiated diamonds; 
         \textbf{b,} High-energy EELS spectra comparing neutron-irradiated and non-irradiated diamonds; 
         \textbf{c,} Low-energy EELS of the same samples; 
         \textbf{d,} Mapping the 285 eV energy loss signal in neutron-irradiated diamond, using a 2 eV selected energy width; 
         \textbf{e,} Energy loss spectra in neutron-irradiated diamond: the red line corresponds to the four square regions in (d) with concentrated sp² bonding states, appearing red in the map, while the blue line represents typical damaged regions, which appear blue in (d). }
\end{figure}
\subsection{Defect clusters and dislocations}
As illustrated in Fig. 1, neutron-irradiated diamond reveals a broad distribution of $\mathrm{sp}^2$ bonding states, which tend to cluster within specific localized regions. Based on this finding, we employed aberration-corrected scanning transmission electron microscopy (STEM) in high-angle annular dark field (HAADF) mode to image irradiation-induced defects with atomic precision from the $\langle 100 \rangle$ orientation in diamond. Fig. 2(a) shows a HAADF atomic-resolution image highlighting the structures in areas where $\mathrm{sp}^2$ bonding states are densely clustered. The irradiation-induced defects visibly disrupt the crystalline lattice structure of diamond. Applying Fourier transform-based geometric phase analysis (GPA) provides a more intuitive visualization of the microstructural alterations within these defect regions. As illustrated in GPA mapping Fig. 2(b), the color gradient represents the computed phase (i.e., distortion). Dislocations are identified by abrupt color changes from blue to red, corresponding to phase discontinuities, as shown in small plot. In $\mathrm{sp}^2$-clustered areas, GPA mapping reveals localized dislocation densities approaching $\sim 10^{14}$ dislocations/$\mathrm{cm}^2$. The presence of these dislocations, induced by irradiation, modifies atomic coordination within the diamond lattice, leading to consequential shifts in bonding structure.
As the distance from the defect cluster increases, the dislocation density decreases.Inverse FFT analysis, filtered with (022) and (0$\overline{22}$) diffraction points, reveals dislocations within the ($\overline{1}$00) plane, as illustrated in Fig. 2(e) and Fig. 2(g). The Burgers vector is found to be less than $1/4\langle 110 \rangle$, which represents the spacing of the closest lattice planes of \{022\} in diamond. In analyzing the relative atomic displacement from the original lattice positions, as depicted in Fig. 2(f) and Fig. 2(h), we identify the corresponding Burgers vectors for the dislocations shown in Fig. 3(e) and Fig. 2(g) as $\mathbf{b}\ =1/6[01\overline{1}]$ and $\mathbf{b}\ =1/12[01\overline{1}]$, respectively. The dislocations are inferred to be produced through the dislocation reaction of two partial dislocations. For instance, through the interaction of two leading Shockley partial dislocations in diamond\cite{Blumenau2003}, $\mathbf{b}_\mathbf{1}=1/6\left[\overline{1}21\right]$ in the $(11\overline{1})$ plane and $\mathbf{b}_\mathbf{1}=1/6\left[1\overline{12}\right]$  in the $(1\overline{1}1)$ plane. Upon interaction, these dislocations combine to form dislocations located in the $(\overline{1}00)$ plane:
\begin{equation}
\frac{1}{6}\left[\overline{1}21\right]+\frac{1}{6}\left[1\overline{12}\right]\rightarrow\frac{1}{6}\left[01\overline{1}\right]
\end{equation}
Or can be rewritten as vectors based of Thompson tetrahedron of diamond in Fig. 2(d):
\begin{equation}
\bm{\gamma B} + \bm{B \alpha} \rightarrow \bm{\gamma \alpha},
\end{equation}
Where $\bm{\gamma \alpha}$ represents a particular configuration of dislocation called Lomer-Cottrell (LC) junction\cite{Lomer1951}\cite{Cottrell1952}. A schematic depiction of the dislocation merging process is provided in Fig. 2(c), illustrating the interactions and resultant configurations. The LC junction dislocations are sessile and immobile in the slip plane, acting as a barrier against other defects in the crystal\cite{Abu-Odeh2022}. This phenomenon can lead to the accumulation of dislocations within the diamond during neutron irradiation, representing a potential source of clustered defect behavior.

Perfect dislocation from HRTEM image in diamond $\langle 110 \rangle$ view with burgers vector $\mathbf{b}\ =1/4\langle110\rangle$ can be seen in Fig. 2(i). The burgers vector is denoted by a yellow arrow and finishes the burgers circuit over the distortion field represented by the white line. The dislocation consists of two edge dislocations with burgers vector $\mathbf{b}\ =1/4[112]$ and $\mathbf{b}\ =1/4[11\overline{2}]$, as denoted inside the white line. The strain field map derived from GPA can be seen in Fig. 2(j), showing the strain $\mathbf{\varepsilon}$ induced by edge dislocations is primarily confined to the immediate vicinity of the dislocation core, influencing an area of approximately nanoscale dimensions. The raw phase of two sets \{111\} planes in Fig. 2(j), shows the configuration of edge-type dislocations. These dislocations possess a Burgers vector of $\mathbf{b}\ =1/4\langle112\rangle$, are commonly observed in diamond cubic as undissociated 60-degree dislocations\cite{Blumenau2003}. A typical example of neighboring dislocations with opposite signs and Burgers vectors of $\mathbf{b}\ =1/4\langle112\rangle$ can be seen in the Supplementary data Fig. S3.

\begin{figure}[H]
\label{fig2}
\centering
\includegraphics[width=\linewidth]{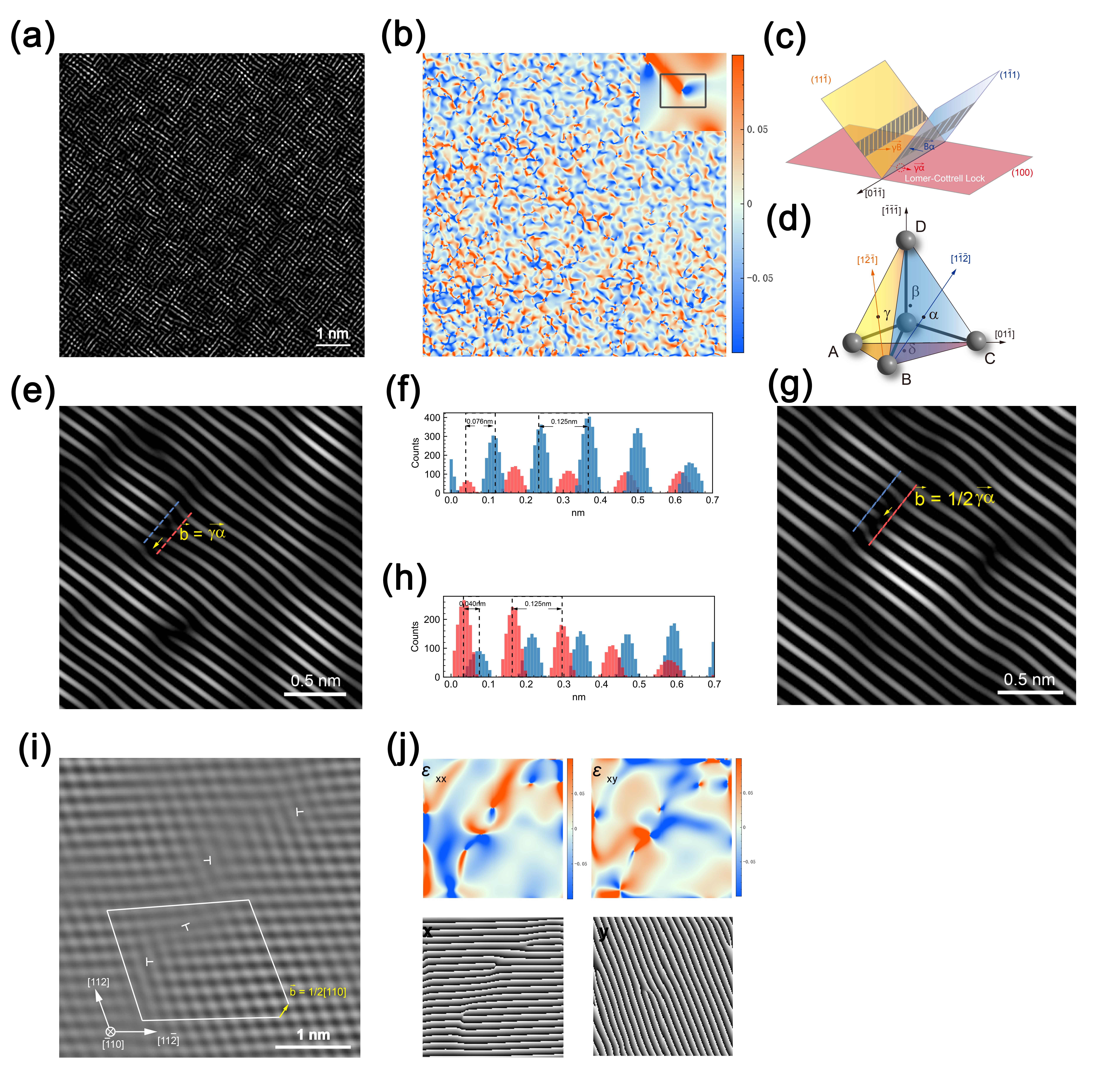}
\caption{TEM image of defect clusters and dislocations in neutron-irradiated sc-CVD diamond. \textbf{a,} FFT filtered HAADF image of defect cluster region in $[\overline{1}00]$ view; 
         \textbf{b,} Strain field of $\mathbf{\varepsilon}_{xy}$ in (a) based on GPA;; 
         \textbf{c,} Schematic diagram of the formation of Lomer-Cottrell lock $\mathbf{b}\ =\bm{\gamma \alpha}$ in \{100\} diamond lattice plane. The reaction process is triggered by the interaction between two leading partial dislocations $\bm{\gamma B}$ in the yellow-colored $(11\overline{1})$ lattice plane and $\bm{B \alpha}$ in the blue-colored $(1\overline{1}1)$ lattice plane; 
         \textbf{d,} Thompson tetrahedron and in diamond lattice and correspond dislocations in (c); 
         \textbf{e,} FFT filtered HAADF image with (022) and $(0\overline{22})$ diffraction points shows Lomer-Cottrell dislocation in $(\overline{1}00)$ diamond lattice plane with a Burgers vector of $\mathbf{b}\ =1/6[01\overline{1}]$, corresponding to $\bm{\gamma \alpha}$ in the Thompson tetrahedron; 
         \textbf{f,} Line profile of atoms around dislocation in (e), blue and red histograms represent the position of the atoms upper and lower the LC dislocation as show lines in (e), the relative displacement between atomic layers is 0.076 nm, approximately 2/3 of interatomic spacing 0.125 nm, which results in $\mathbf{b} =1/6\langle110\rangle$;
         \textbf{g,} Lomer-Cottrell dislocation with a Burgers vector of $\mathbf{b}\ =1/12[01\overline{1}]$ corresponding to $1/2 \bm{\gamma \alpha}$ in the Thompson tetrahedron; 
         \textbf{h,} Line profile of atoms around dislocation in (g), similarly, the relative displacement between atomic layers is 0.04 nm, approximately 1/3 of interatomic spacing corresponds to $\mathbf{b} =1/12\langle110\rangle$; 
         \textbf{i,} HRTEM image shows dislocation with $\mathbf{b}\ =1/2[110]$ in $[\overline{1}10]$ view, and two opposite signs with $\mathbf{b}\ =1/4[112]$ and $\mathbf{b}\ =1/4[\overline{112}]$ shown in the upper part of the image;
\textbf{i,}Strain field $\mathbf{\varepsilon}_{xx}$, $\mathbf{\varepsilon}_{xy}$, and raw phase in x,y direction from GPA methods.}
\end{figure}

\subsection{Carbon phase transition}
Under neutron irradiation, localized carbon phase transformations in diamond can be observed in the HRTEM image along the $\langle 110 \rangle$ zone axis, shown in Fig. 3(a). Four regions are marked: the diamond structure in blue and potential graphitic structures in green, red, and purple. FFT patterns from these regions, displayed in Fig. 3(b), reveal that the blue region’s FFT closely resembles that of diamond, while the FFT images of the three marked graphitic regions show significant deviations, indicating a shift toward a graphitic structure. Fig. 3(c) zooms into the upper-right section of Fig. 3(a), revealing the characteristic {111} planes of diamond with a lattice spacing of 0.205 nm in the blue boxed region, while the red box displays a hexagonal pattern, indicative of graphitic features. Further enlargement of the red box in Fig. 3(c), shown in Fig. 3(d), highlights a structure resembling graphitic layers viewed along the [223] orientation, which matches the experimentally observed configuration. To verify this hypothesis, we simulated HRTEM imaging of graphite, aligning the electron beam along the [223] orientation. This setup places the graphite [001] axis at an 11.27° angle to the electron beam. The simulated image, is presented in Fig. 3(g), aligns closely with the periodic structural intensity profiles observed in the experimental HRTEM data, as shown in Fig. 3(f). Specifically, the blue profile represents the simulation results (Fig. 3(g)), while the red profile corresponds to the experimental measurements (Fig. 3(d)). FFT comparisons for both experimental and simulated results are shown in Fig. S3(e) and S3(f), respectively. This structural and imaging comparison confirms the emergence of graphitic phases within neutron-irradiated diamond. These graphitic domains, embedded at a nanometric scale within the diamond lattice, represent an additional source of $\mathrm{sp}^2$-bonded carbon.
\begin{figure}[H]
\label{fig3}
\centering
\includegraphics[width=0.9\linewidth]{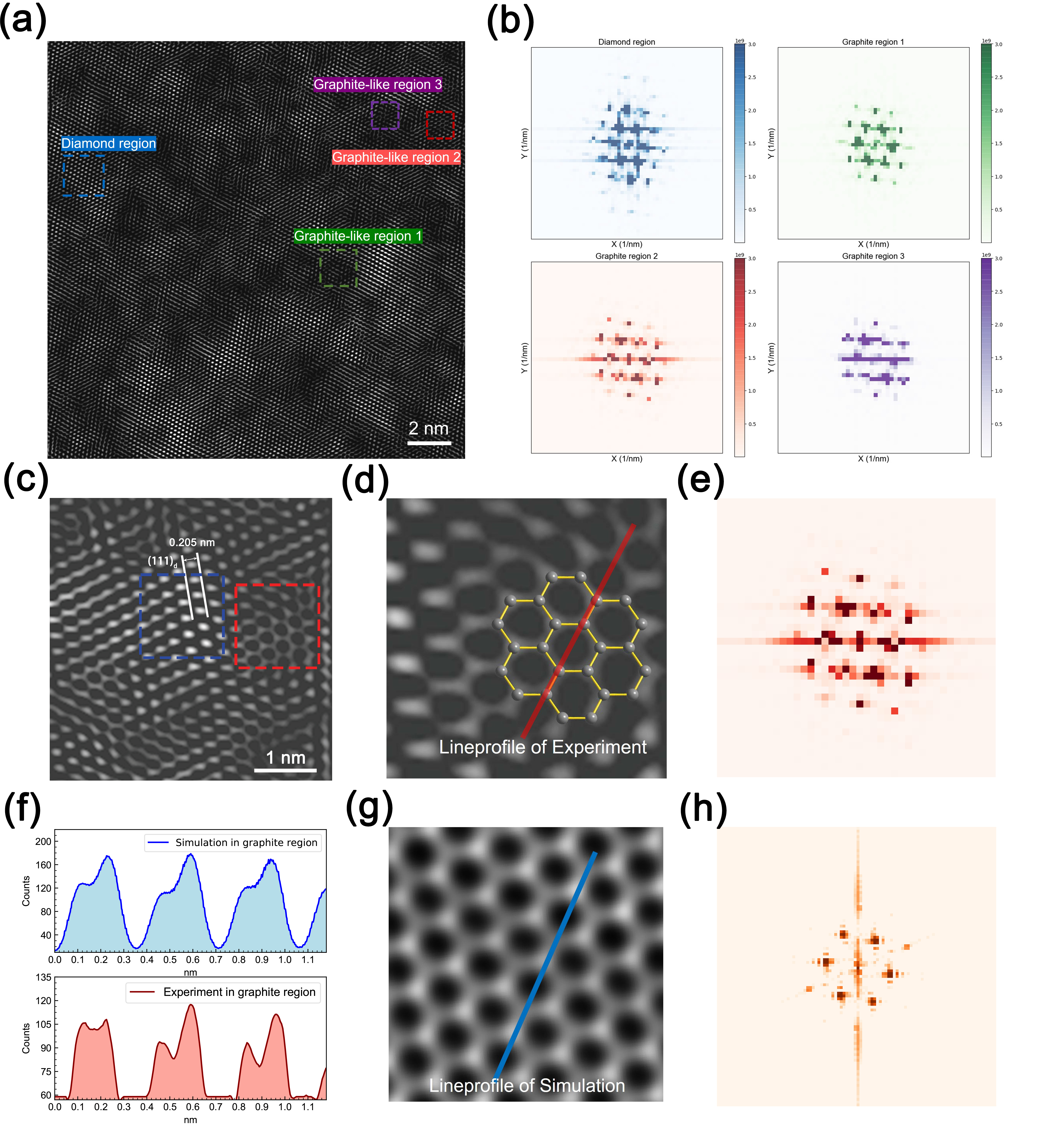}
\caption{HRTEM image illustrating carbon phase transition in neutron-irradiated diamond. 
\textbf{a,} HRTEM image of neutron-irradiated diamond along the ⟨110⟩ crystallographic direction, processed with FFT filtering to highlight structural details; 
         \textbf{b,} FFT patterns from selected regions in the irradiated diamond, showing distinct phases where the diamond phase is marked in blue, and graphite-like regions are represented in green, red, and purple as indicated in (a); 
         \textbf{c,} Magnified view of a selected area from (a), showing diamond and graphite-like carbon phase in the same region. The blue box highlights the fringes of the diamond (111) plane with a spacing of 0.205 Å, while the red box indicates a graphite-like structure; 
         \textbf{d,} Magnified image from the red box in (c), showing a possible graphite layer structure with an angle of 11.27°relative to the electron beam direction; 
         \textbf{e,} FFT of the graphite-like region captured in the HRTEM experiment (d); 
         \textbf{f,} Line profile of the graphite-like region in HRTEM simulation and experiment, with periodicity blue histogram representing the blue line in the HRETM simulation (g), and periodicity red histogram representing the red line in the HRTEM experiment (d);
         \textbf{g,} Simulated HRTEM image with the graphite [001] orientation tilted at an angle of 11.27° relative to the beam incident direction, actually representing imaging along the graphite $[2\overline{2}3]$ zone axis; (h) The FFT image of the graphite-like simulated HRTEM image in (g) shows a pattern highly similar to the experimental FFT pattern in (e).}
\end{figure}

\subsection{Imaging the dumbbell-like interstitials and analysis point defect hidden in diamond lattice}
Radiation-induced dumbbell interstitial and vacancy defects are directly visualized in the HAADF image depicted in Fig. 4 (a). The interstitial in $[\overline{1}00]$ view, highlighted by the white dashed boxes in Fig. 4 (a) denoted by "i", Fig. 4 (b), and Fig. 4 (c), appears as dumbbell configurations where two atoms share a single lattice site. This configuration, widely recognized in the literature to form under irradiation in diamond\cite{Zhang2018}, is directly observed in this work for the first time. The alignment of two interstitial atoms determines the specific configuration of each dumbbell defect. In Fig. 4(a) and Fig. 4(c), the projections of dumbbell orientations within the (100) plane align along the ⟨011⟩ direction, whereas in Fig. 4(b), the projection lies along the ⟨001⟩ direction. From diamond's $[\overline{1}10]$ orientation view, as shown in the HRTEM images Fig. 4(d) and Fig. 4(e), Fig. 4(d) illustrates a dumbbell configuration projecting within the $(\overline{1}10)$ plane along the ⟨001⟩ direction, while Fig. 4(e) displays an alternative dumbbell projection within the $(\overline{1}10)$ plane along the ⟨111⟩ direction, leading to distinct contrast variations. Combining these projection configurations across orientations, it is evident that irradiation-induced dumbbell structures preferentially align along the ⟨100⟩ and ⟨111⟩ directions. This finding substantiates that dumbbell interstitial defects in diamond can form stable states along the ⟨100⟩ and ⟨111⟩ orientations, while the ⟨110⟩ direction remains unstable according to theoretical calculations\cite{Saada1998}\cite{Breuer1995}. 

A vacancy defect is also marked by a white dashed box and labeled 'v' in Fig.4 (a), presents a reduced imaging intensity within the atomic column, as shown in the line profile from the top-left corner. In addition, along with the intensity divergence of each atom column visualized in jet color from HAADF results, atomic columns with similar intensities are found to be distributed close to one another based on the intensity information depicted in the color map, indicating the tendency to aggregate the point defects. This clustering behavior of point defects may eventually lead to dislocation formation\cite{was2016fundamentals}. The interconnected double dumbbell structures are depicted in Fig. 4(c), provide further evidence of these processes and suggest early-stage development of dislocations, such as Frank dislocation loops. Three dumbbell-shaped interstitials along the ⟨100⟩ direction are depicted in Fig. 4(f), Fig. 4(g), and Fig. 4(h). The dumbbell structure, aligned with the ⟨100⟩ direction, is confirmed by TEM and provides a basis for further investigation into the correlation between irradiation-induced defects and spectroscopic features. The line profile in Fig. 4(c) and additional vacancy-type point defects are provided in Fig. S1 and Fig. S2 of the Supplementary data.

Raman and EELS analyses indicate that a large portion of the diamond lattice retains its order after irradiation, as also observed in TEM images. Fig. 5(b) and Fig. 5(e) show neutron-irradiated imaging along the diamond ⟨100⟩ and ⟨110⟩ directions, respectively. Visually, these irradiated lattice structures appear similar to the non-irradiated diamond images in Fig. 5(a) and Fig. 5(d), with the main difference being a looser atomic distribution and slightly larger lattice spacing. The diffraction patterns in the upper-right corners of Fig. 5(b) and Fig. 5(e) confirm that no extensive new structure has formed within the diamond due to irradiation. However, point defects, such as interstitials and vacancies, may be concealed within the atomic columns along the electron beam path, making them challenging to detect directly. Defects such as local point defects or dislocations not resolved by TEM contrast (e.g., due to Burgers vectors aligned along the observation direction) may still be present.

\begin{figure}[H]
\label{fig4}
\centering
\includegraphics[width=0.9\linewidth]{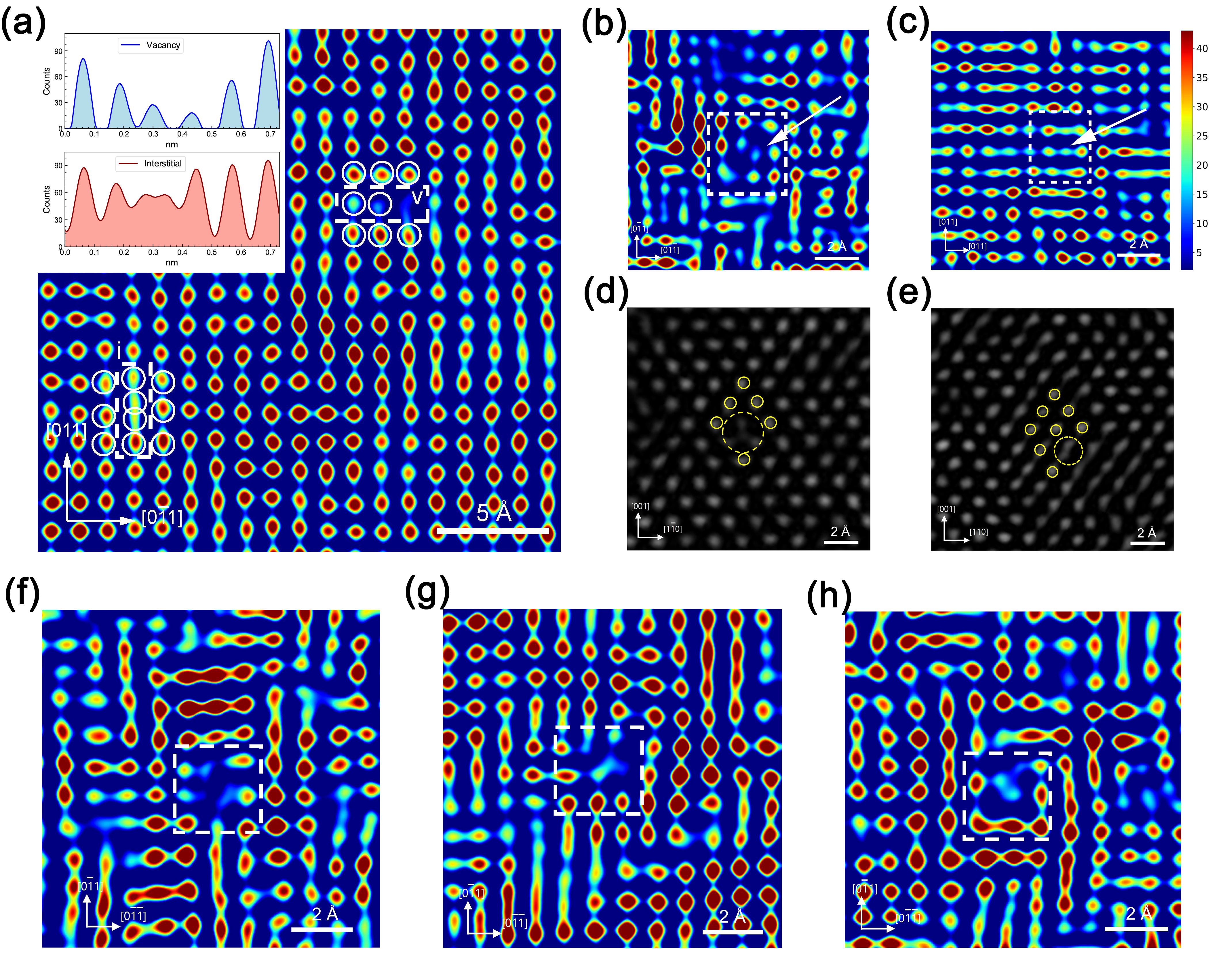}
\caption{Direct TEM imaging of point defects in neutron-irradiated diamond. 
\textbf{a,} Interstitial and vacancy defects are indicated with white dashed boxes in the FFT filtered HAADF image from ⟨100⟩ electron beam direction, marked ‘ i ’ and ‘ v ’ respectively. The line profile along the long edge of the box (top left corner of the image) reflects the corresponding atomic structures; 
         \textbf{b,} FFT filtered HAADF image of ⟨100⟩ dumbbell interstitial in diamond from [100] view; 
         \textbf{c,} FFT filtered HAADF image of ⟨111⟩ dumbbell interstitial from[100] view; 
         \textbf{d,} FFT filtered HRTEM image of ⟨100⟩ dumbbell interstitial from [110] view; 
         \textbf{e,} FFT filtered HRTEM image of ⟨111⟩ dumbbell interstitial from [110] view; 
         \textbf{f$\sim$h,} FFT filtered HAADF image of ⟨100⟩ dumbbell interstitial from [100] view.}
\end{figure}

To quantitatively assess hidden point defects and distortions in the observed lattice, we conducted a comparative analysis of the lattice spacing and atomic number density within each atom column between neutron-irradiated and non-irradiated diamond. Atomic coordinates for the images in Fig. 5(a), Fig. 5(b), Fig. 5(d), and Fig. 5(e) were determined using a two-dimensional Gaussian fitting algorithm\cite{DeBacker2016}\cite{VanAert2009}. Interatomic distances derived from these coordinates were then compiled into histograms. Fig. 5(c) and Fig. 5(f) illustrate the statistical distribution of atomic spacings in the ⟨110⟩ direction in diamond ⟨100⟩ view and in the ⟨111⟩ direction in diamond ⟨110⟩ view for both irradiated and non-irradiated diamond.
In the HAADF images, the detector captures signals from high-angle scattered electrons, with each atomic column’s intensity $I_s$ representing the number of scattered electrons, which is proportional to the number of atoms in each column, as described by Pennycook\cite{Pennycook1988}\cite{Pennycook1990}\cite{Pennycook1991}: $I_s=\sigma_sNtI_{in}$, where $I_s$ is the intensity of scattered electrons (corresponding to the HAADF image intensity), $I_{in}$ is the incident electron beam intensity, $N$ is the number of atoms in each column, $\sigma_s$ is the scattering cross-section of electrons with carbon atoms in diamond, and $t$ is the sample thickness.
Using the two-dimensional Gaussian fit, the intensity $I_s$ for each column was calculated by summing the pixel values under the Gaussian peak in the image, reflecting the atom count per column. As illustrated in Fig. 5(g), the intensity values were also compiled into histograms to compare the atomic column population distribution in neutron-irradiated and non-irradiated diamond.

\begin{figure}[H]
\label{fig5}
\centering
\includegraphics[width=0.8\linewidth]{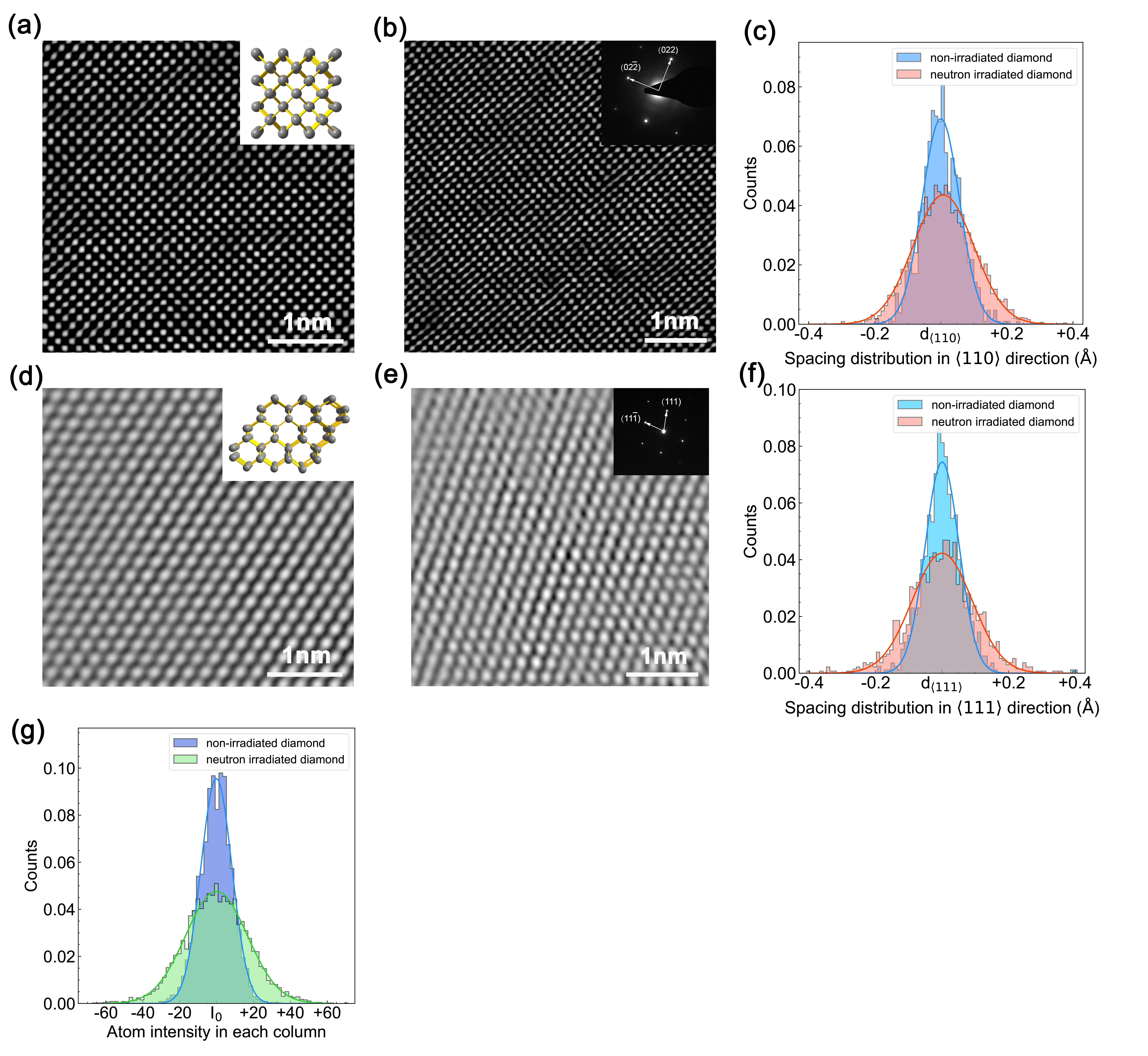}
\caption{Comparison of long-range ordered lattice regions in neutron-irradiated and non-irradiated sc-CVD diamonds. 
\textbf{a,} HAADF image of $[\overline{1}00]$ lattice view of non-irradiated diamond with atom model in the upper right corner; 
         \textbf{b,}  HAADF image of $[\overline{1}00]$ lattice view of neutron-irradiated sc-CVD diamond; 
         \textbf{c,} Distribution of lattice spacing along the ⟨110⟩ direction in $[\overline{1}00]$ view, the red histogram represents the lattice spacing in neutron-irradiated diamond, while the blue histogram shows the lattice spacing in non-irradiated diamond; 
         \textbf{d,} HRTEM image of $[\overline{1}10]$ lattice view of non-irradiated diamond; 
         \textbf{e,} HRTEM image of $[\overline{1}10]$ lattice view of neutron-irradiated diamond; 
         \textbf{f,} Distribution of lattice spacing along the ⟨111⟩ direction in $[\overline{1}10]$ view, the red histogram represents the lattice spacing in neutron-irradiated diamond, and the blue histogram represents that in non-irradiated diamond;
         \textbf{g,} Distribution of atomic intensity in each atomic column in the $[\overline{1}00]$ view (HAADF mode), illustrating the presence of point defects within the atomic columns. The green histogram represents the density distribution of atoms per column in neutron-irradiated diamond, while the blue histogram represents that in non-irradiated diamond.}
\end{figure}
Fitting the histograms of irradiated and non-irradiated samples reveals numerous hidden vacancies and interstitial-type point defects in the irradiated diamond. These defects are challenging to observe directly at the atomic scale in TEM images but manifest as irregularities in the atomic spacing distribution and increased variance in atomic column intensity across different crystal orientations. The standard deviation $S=\sqrt{E(\mathbf{X}-\mu)^2}$ of the atomic spacing and column intensity distributions quantifies the deviation from theoretical atomic distances after irradiation. The standard deviation values are provided in Table 1.

\begin{table}[H]
\label{table1}
\centering
\caption{Statistical deviations density functions of atomic spacing and atomic column density in diamond before and after neutron irradiation} \par
\begin{tabular}{ccc}
\toprule
Measurement of Statistical deviations &non-irradiated & neutron-irradiated  \\ 
\midrule
Atomic spacing on \{100\} plane (Å) & 0.056&0.09 \\ 
Atomic spacing on \{110\} plane (Å) & 0.051& 0.09\\ 
Atom numbers in each column along [100] & 8.4& 16.7\\ 
\bottomrule
\end{tabular}
\end{table}
The nearly doubled standard deviation of atomic spacings in irradiated diamond indicates substantial lattice distortions, due to local point defects or undetected dislocations not resolved by TEM contrast. Additionally, the increased standard deviation of atom number density in each column suggests that, although the neutron-irradiated diamond largely maintains its long-range periodic structure as seen in Fig. 5, the atomic density within each column fluctuates noticeably, indicating the widespread presence of vacancies or self-interstitials throughout the crystal.

\subsection{Simulation of primary and secondary damage in diamond}
Spectroscopic and TEM analyses reveal that neutron irradiation damage induces a non-uniform defect distribution. Some areas contain dense clusters of defects, while others show only isolated point defects located between atomic columns. For the characterization results of fast neutron radiation damage in diamond, we performed a simulation combining Monte Carlo (MC) and molecular dynamics (MD) methods to reveal its generation process, corresponding to real radiation experiments. The energy spectrum distribution of fast neutron-induced primary knock-on atoms (PKAs) was obtained with MC, whereby a recoiled diamond atom is defined as a PKA if its kinetic energy exceeds its displacement energy, as illustrated in Fig. 6(a). The observed energy distribution of PKAs ranges from 0 to 280 eV, indicative of the neutron-nuclear elastic scattering process underlying primary damage. The penetration length of these energetic PKAs inside the diamond was also determined from MC simulations, as shown in Fig. 6(a). Within a range of 240 nm, they rapidly deposit all their energy, thus initiating a cascade process through interactions with carbon atoms in the diamond lattice, ultimately becoming the key factor behind crystal damage. Selecting a region of diamond crystal measuring 100 × 100 × 100 $\mathrm{nm3}$, Fig. 6(b) illustrates the spatial distribution of PKAs in this region, showing a total of 199 PKAs generated after neutron irradiation at a dose of $3\times10^{17}$n/$\mathrm{cm}^2$. Molecular dynamics was employed to investigate the dynamic interactions between energetic PKA particles and diamond lattice. A diamond lattice comprising 1,000,000 atoms arranged in a 50 × 50 × 50 unit cell configuration was constructed, and PKA particles with energies of 1, 3, 5, 7, 12, and 20 keV were selected for the simulations. Evaluation of radiation damage was based on the production of Frenkel pairs, as illustrated in Fig. 6(c). As the energy of PKAs increased, a proportional rise in the generation of displacement atoms within the diamond was observed, as illustrated in Fig. 6(d). According to MC simulation results, an average of 1.1 PKA particles are produced within a single MD simulation box during realistic irradiation experiments. Hence it can be inferred that the multitude of defects observed in neutron-irradiated diamond reflects the cumulative effects of continuously generated PKAs undergoing cascading processes. As an example, the lattice damage and cascade process caused by a 7 keV incident PKA in diamond, resulting in interstitials and vacancies, can be visualized in the Supplementary data Fig. S4 and Fig. S5.

\begin{figure}[H]
\label{fig6}
\centering
\includegraphics[width=\linewidth]{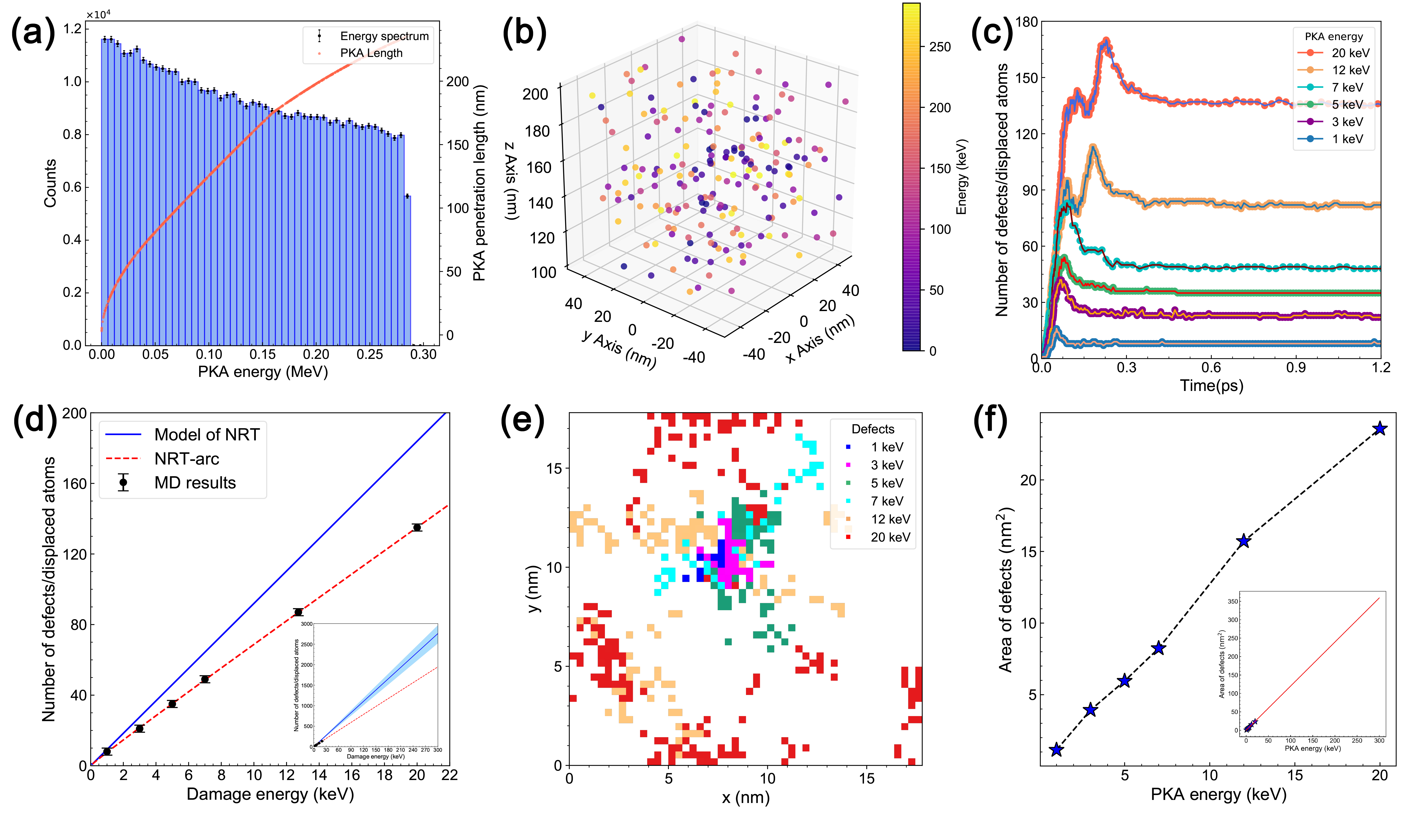}
\caption{Simulated primary damage caused by PKAs in neutron-irradiated sc-CVD diamond and the resulting secondary defect damage to the lattice. 
\textbf{a,} HAADF image of $[\overline{1}00]$ lattice view of non-irradiated diamond with atom model in the upper right corner; 
         \textbf{b,}  HAADF image of $[\overline{1}00]$ lattice view of neutron-irradiated sc-CVD diamond; 
         \textbf{c,} Distribution of lattice spacing along the ⟨110⟩ direction in $[\overline{1}00]$ view, the red histogram represents the lattice spacing in neutron-irradiated diamond, while the blue histogram shows the lattice spacing in non-irradiated diamond; 
         \textbf{d,} HRTEM image of $[\overline{1}10]$ lattice view of non-irradiated diamond; 
         \textbf{e,} HRTEM image of $[\overline{1}10]$ lattice view of neutron-irradiated diamond; 
         \textbf{f,} Distribution of lattice spacing along the ⟨111⟩ direction in $[\overline{1}10]$ view, the red histogram represents the lattice spacing in neutron-irradiated diamond, and the blue histogram represents that in non-irradiated diamond;
         \textbf{g,} Distribution of atomic intensity in each atomic column in the $[\overline{1}00]$ view (HAADF mode), illustrating the presence of point defects within the atomic columns. The green histogram represents the density distribution of atoms per column in neutron-irradiated diamond, while the blue histogram represents that in non-irradiated diamond.}
\end{figure}

To quantitatively estimate the damage caused to the crystal by PKAs with different energies, the Norgett-Robinson-Torrens (NRT) model\cite{Robinson1974}\cite{Nordlund2018NC} corrected by the adiabatic recombination process (NRT-arc)\cite{Nordlund2018NC} was employed to assess the number of defects $N_{\mathrm{d},arc}$ produced by PKAs with varying energies $T_d$ on diamond:
\begin{equation}
N_{\mathrm{d},arc}\left(T_\mathrm{d}\right)=\frac{0.8T_\mathrm{d}}{2E_\mathrm{d}}\cdot\xi_{arc}\left(T_\mathrm{d}\right)\mathrm{\ \ \ (for\ \ \ }T_\mathrm{d}\geq2E_\mathrm{d}/0.8)
\end{equation}
where $E_d$ is diamond displacement energy, $\xi_{arc}$ is an effective function that represents the adiabatic recombination process, which can be expressed as 
\begin{equation}
\xi_{arc}\left(T_\mathrm{d}\right)=\frac{1-c_{arc}}{\left(2E_\mathrm{d}/0.8\right)^{b_{arc}}}\cdot T_\mathrm{d}^{b_{arc}}+c_{arc}
\end{equation}

The exponent $b_{arc}$ characterizes how the PKA range depends on energy, and the unitless quantity $c_{arc}$ relates to the saturation value of damage recombination concerning heat spike size. Using the number of stable displacements from MD as input parameters, we fitted the parameters of the NRT-arc model, obtaining $b_{arc}$ = -0.369 ± 0.064 and $c_{arc}$ = 0.689 ± 0.019, as shown in Fig. 6(d). These parameters enable us to evaluate the damage caused by different energy PKAs in the radiated diamond materials. With increasing PKA energy and quantity, various types of dislocations also begin to appear, over time and scale, these dislocations can undergo increasingly complex dynamic processes. Different directions of Shockley-type dislocations have a significant probability of forming LC locks, further enhancing the aggregation of dislocations and the formation of clusters. Fig. 6(e) illustrates the projection of defective regions within the diamond onto a plane perpendicular to the direction of PKA incidence, where different color blocks correspond to defects induced by PKAs of varying energy levels. The defect area increases with the energy of the incident particles, as evidenced by the trend observed in Fig. 6(f). It can be anticipated that PKAs induced by fast neutrons can cascade to generate damage extending up to several hundred nanometers in range. Combining MD and MC simulation methods allows us to compare the damage caused by neutrons and recoil carbon atoms. Based on the incident particle flux and the number of PKAs obtained from MC simulations, we find that on average, one fast neutron generates 0.027 PKAs in a diamond thin film crystal. In contrast, according to the NRT-arc model, a 140 keV PKA can produce six thousand Frenkel pairs. From the perspective of cascade damage, these PKAs exhibit a damage capacity in the diamond crystal that is one million times greater than that of fast neutrons. It can be foreseen that clusters of point defects and dislocations are introduced across a large-area crystal during cascade processes and the accumulation of point defects, potentially leading to localized amorphization or even phase transitions.

Through analysis of the damaged lattice structure from simulation, stable structures of dumbbell interstitials along the ⟨111⟩ and ⟨100⟩ directions induced by recoil atoms were observed, as illustrated in Fig. 7(a)-(d). By simulating the imaging of these structures in HAADF and HRTEM mode, as depicted in Fig. 7(e)-(h), these observations are consistent with the TEM experimental findings Fig. 7(i)-(l). The ⟨100⟩ dumbbell interstitial defect exhibits distinct contrast features depending on the crystal orientation. When observed along the $[\overline{1}00]$ direction, the defect reveals the contrast of two interstitial atoms aligned along the ⟨100⟩ direction. In the $[\overline{1}10]$ orientation, the contrast indicates an additional atom connecting the neighboring three atomic columns. For the ⟨111⟩ dumbbell interstitial defect, the $[\overline{1}00]$ crystal viewing reveals the insertion of an atom into the diamond's ⟨110⟩ atomic arrangement. In the $[\overline{1}10]$ orientation, the contrast highlights interstitial atoms within the (111) atomic plane, appearing as continuous contrast along the atomic direction. 

\begin{figure}[H]
\label{fig7}
\centering
\includegraphics[width=\linewidth]{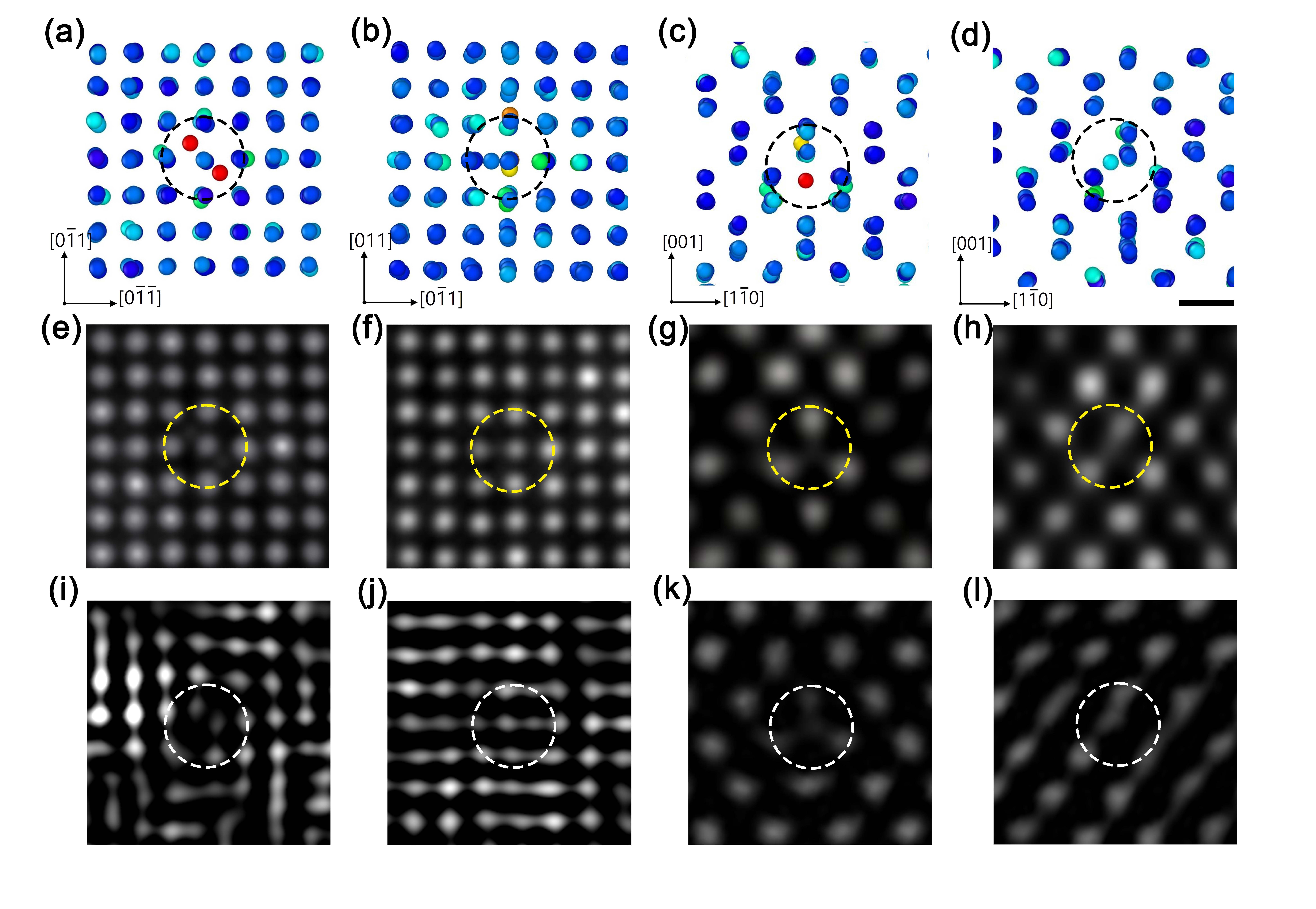}
\caption{TEM direct imaging of ⟨100⟩ and ⟨111⟩ dumbbell interstitial defects in neutron-irradiated diamond, alongside molecular dynamics simulation images and corresponding TEM imaging simulations based on molecular dynamics structures. 
\textbf{a$\sim$d,} Atomic structures of dumbbell interstitial defects generated by PKA incidence in diamond, obtained from molecular dynamics simulations; 
         \textbf{e$\sim$h,}  Simulated HAADF and HRTEM images based on molecular dynamics structures of the dumbbell interstitial defects, corresponding to (a)$\sim$(d); 
         \textbf{i and j,} HAADF images along the $[\overline{1}00]$ zone axis, showing dumbbell-shaped interstitial defects oriented along the ⟨100⟩ and ⟨111⟩ directions; 
         \textbf{k and l,} HRTEM images along the $[\overline{1}10]$ zone axis, capturing ⟨100⟩ and ⟨111⟩ dumbbell interstitial defect structures.}
\end{figure}

Neutron irradiation offered an excellent opportunity to investigate the behavior of primary defects thoroughly due to its penetrating nature. Typically studies of radiation damage in the crystal structure of diamond have relied on spectroscopic methods. As previously mentioned, self-split interstitials generated during irradiation have been investigated using various spectroscopic techniques, such as Raman spectroscopy, where the possible presence of the 1630 $\mathrm{cm}^{-1}$ band has been studied \cite{orwa_raman_2000}\cite{khomich_features_2019}, photoluminescence (PL) spectroscopy, which reveals the presence of 3H and TR12 centers \cite{walker_optical_1977}, and electron paramagnetic resonance (EPR), where they manifest as the characteristic R2 defect in type-IIa diamonds subjected to electron irradiation \cite{prb67}. This work provides direct imaging of ⟨100⟩ and ⟨111⟩ dumbbell-like self-split interstitials, offering experimental support for constructing defect models. The structure of two ⟨111⟩ dumbbell-like split-interstitials can be viewed as the origin of interstitial complexes, potentially marking the onset of dislocation formation. Irradiation triggers a transition from $\mathrm{sp}^3$ to $\mathrm{sp}^2$ hybridization in diamond, resulting in a 3\% $\mathrm{sp}^2$ content observed in EELS. Experimental results reveal that immobile dislocations, such as Lomer-Cottrell dislocations, can serve as nucleation centers for defect clusters during irradiation. These clusters exhibit a $\mathrm{sp}^2$ content five times higher than other defect regions, likely leading to the formation of different carbon structures. From a defect formation dynamics perspective, comparing point defect quantities estimated using the NRT-arc method allows for assessing crystal damage in diamond under varying irradiation times, which is pivotal for materials science and the application of diamond-based devices in harsh environments. Energetic recoil atoms play a crucial role in localized lattice distortion. We can conveniently assess damage levels by comparing PKA spectra from different radiations and estimating NRT-arc, assuming defect interactions are not excessively severe.

\section{Conclusion}
The degradation of the performance of diamond devices in irradiation environments has long been attributed to the formation of internal defects that increase carrier scattering. This study provides a detailed imaging analysis of defects in diamond under high-dose neutron irradiation, offering insights into their composition and structure. Radiation-induced defects are identified in Raman spectra as interstitial structures, which cause local $\mathrm{sp}^3$ to $\mathrm{sp}^2$ bonding transitions. Through EELS, the aggregation tendency of sp² structures is revealed, highlighting the formation of defect clusters. TEM characterization further enriches these findings. From a morphological perspective, it unveils the defect content within clusters, identifies LC-lock type dislocations induced by irradiation, and deciphers the atomic structures of two distinct dumbbell like interstitial configurations. In terms of distribution, TEM reveals different regimes, ranging from densely clustered defects with localized phase transformations, indicative of advanced stages of defect evolution, to sparsely distributed point defects, representing the initial stages of irradiation-induced damage. Dynamic simulations demonstrate that defect formation begins with point defects, which condense into dislocations and clusters as interstitial atoms gradually aggregate. The structural phase transformation of carbon evolves in parallel with the progression of defect formation. This study provides an experimental insight into the atomic-scale structural changes induced by irradiation in diamond, advancing our understanding of radiation damage effects and mechanisms in diamond, and offer valuable references for the design and optimization of diamond-based devices for high-radiation applications with irradiation damage studies.

\section*{Acknowledgment}
The authors would like to acknowledge M. Bulavin, A. Cheplakov, L. Kurchaninov, V. Kukhtin, J. Ye, C. Li, H. Ding, J. Du, Z. Zou and F. Miao for their testing facilities and helpful discussions. They would also like to acknowledge M. Aleksa, A. Straessner, H. Kagan, G. Oakham, W. Trischuk, J. Rutherfoord, R. McPherson and the ATLAS-LAr Collaboration for their strong supports and beneficial comments. This work was supported by the“International Science \& Technology Cooperation Program of China”(Contract No. 2015DFG02100), The Ministry of Science and Technology of the People's Republic of China.

\section*{Data availability}
Data will be made available on request.

\section*{Declaration of interests}
The authors declare that they have no known competing financial interests or personal relationships that could have appeared to influence the work reported in this paper.
\section*{CRediT authorship contribution statement}
\textbf{Jialiang Zhang:} Conceptualization, Methodology, Investigation,Formal Analysis, Software, Writing – original draft;
\textbf{Futao Huang:} Methodology, Investigation, Formal Analysis, Software;
\textbf{Shuo Li:} Formal Analysis, Writing – review \& editing;
\textbf{Guojun Yu:} Formal Analysis, Writing – review \& editing;
\textbf{Zifeng Xu:} Formal Analysis, Writing – review \& editing;
\textbf{Lifu Hei:} Resources,Writing – review \& editing;
\textbf{Fanxiu Lv:} Resources,Writing – review \& editing;
\textbf{Aidan Horne:} Methodology, Investigation Formal Analysis, Software;
\textbf{Peng Wang:} Conceptualization, Methodology, Writing – review \& editing,Supervision;
\textbf{Ming Qi:} Conceptualization, Methodology, Funding acquisition, Writing – review \& editing, Supervision;

\bibliographystyle{elsarticle-num-names}
\bibliography{output}%

\begin{thebibliography}{67}
\expandafter\ifx\csname natexlab\endcsname\relax\def\natexlab#1{#1}\fi
\providecommand{\url}[1]{\texttt{#1}}
\providecommand{\href}[2]{#2}
\providecommand{\path}[1]{#1}
\providecommand{\DOIprefix}{doi:}
\providecommand{\ArXivprefix}{arXiv:}
\providecommand{\URLprefix}{URL: }
\providecommand{\Pubmedprefix}{pmid:}
\providecommand{\doi}[1]{\href{http://dx.doi.org/#1}{\path{#1}}}
\providecommand{\Pubmed}[1]{\href{pmid:#1}{\path{#1}}}
\providecommand{\bibinfo}[2]{#2}
\ifx\xfnm\relax \def\xfnm[#1]{\unskip,\space#1}\fi
\bibitem[{Field(2012)}]{Field2012}
\bibinfo{author}{J.~E. Field},
\newblock \bibinfo{title}{{The mechanical and strength properties of diamond}},
\newblock \bibinfo{journal}{Reports on Progress in Physics}
  \bibinfo{volume}{75} (\bibinfo{year}{2012}).
  \DOIprefix\doi{10.1088/0034-4885/75/12/126505}.
\bibitem[{Childress et~al.(2006)Childress, {Gurudev Dutt}, Taylor, Zibrov,
  Jelezko, Wrachtrup, Hemmer, and Lukin}]{Childress2006}
\bibinfo{author}{L.~Childress}, \bibinfo{author}{M.~V. {Gurudev Dutt}},
  \bibinfo{author}{J.~M. Taylor}, \bibinfo{author}{A.~S. Zibrov},
  \bibinfo{author}{F.~Jelezko}, \bibinfo{author}{J.~Wrachtrup}, et~al.,
\newblock \bibinfo{title}{{Coherent Dynamics of Coupled Electron and Nuclear
  Spin Qubits in Diamond}},
\newblock \bibinfo{journal}{Science}  \bibinfo{volume}{314}
  (\bibinfo{year}{2006}) \bibinfo{pages}{281--285}.
  \DOIprefix\doi{10.1126/science.1131871}.
\bibitem[{Tapper(2000)}]{Tapper2000}
\bibinfo{author}{R.~J. Tapper},
\newblock \bibinfo{title}{{Diamond detectors in particle physics}},
\newblock \bibinfo{journal}{Reports on Progress in Physics}
  \bibinfo{volume}{63} (\bibinfo{year}{2000}) \bibinfo{pages}{1273--1316}.
  \DOIprefix\doi{10.1088/0034-4885/63/8/203}.
\bibitem[{Narayan et~al.(2016)Narayan, Jones, Cornejo, Dalton, Deconinck,
  Dutta, Gaskell, Martin, Paschke, Tvaskis, Asaturyan, Benesch, Cates, Cavness,
  Dillon-Townes, Hays, Ihloff, Jones, King, Kowalski, Kurchaninov, Lee,
  McCreary, McDonald, Micherdzinska, Mkrtchyan, Mkrtchyan, Nelyubin, Page,
  Ramsay, Solvignon, Storey, Tobias, Urban, Vidal, Waidyawansa, Wang, and
  Zhamkotchyan}]{Narayan2016}
\bibinfo{author}{A.~Narayan}, \bibinfo{author}{D.~Jones},
  \bibinfo{author}{J.~C. Cornejo}, \bibinfo{author}{M.~M. Dalton},
  \bibinfo{author}{W.~Deconinck}, \bibinfo{author}{D.~Dutta}, et~al.,
\newblock \bibinfo{title}{{Precision electron-beam polarimetry at 1 GeV using
  diamond microstrip detectors}},
\newblock \bibinfo{journal}{Physical Review X}  \bibinfo{volume}{6}
  (\bibinfo{year}{2016}) \bibinfo{pages}{1--9}.
  \DOIprefix\doi{10.1103/PhysRevX.6.011013}.
\bibitem[{Pace et~al.(2000)Pace, {Di Benedetto}, and Scuderi}]{Pace2000}
\bibinfo{author}{E.~Pace}, \bibinfo{author}{R.~{Di Benedetto}},
  \bibinfo{author}{S.~Scuderi},
\newblock \bibinfo{title}{{Fast stable visible-blind and highly sensitive CVD
  diamond UV photodetectors for laboratory and space applications}},
\newblock \bibinfo{journal}{Diamond and Related Materials}  \bibinfo{volume}{9}
  (\bibinfo{year}{2000}) \bibinfo{pages}{987--993}.
  \DOIprefix\doi{10.1016/S0925-9635(00)00213-2}.
\bibitem[{Ogawa et~al.(2023)Ogawa, Isobe, Weiss, Griesmayer, Sangaroon, Takada,
  Masuzaki, Ohtani, Liao, Tamaki, Murata, and Osakabe}]{Ogawa2023}
\bibinfo{author}{K.~Ogawa}, \bibinfo{author}{M.~Isobe},
  \bibinfo{author}{C.~Weiss}, \bibinfo{author}{E.~Griesmayer},
  \bibinfo{author}{S.~Sangaroon}, \bibinfo{author}{E.~Takada}, et~al.,
\newblock \bibinfo{title}{{Fusion product diagnostics based on commercially
  available chemical vapor deposition diamond detector in large helical
  device}},
\newblock \bibinfo{journal}{Journal of Instrumentation}  \bibinfo{volume}{18}
  (\bibinfo{year}{2023}) \bibinfo{pages}{P01022}.
  \DOIprefix\doi{10.1088/1748-0221/18/01/P01022}.
\bibitem[{Wort and Balmer(2008)}]{Wort2008}
\bibinfo{author}{C.~J. Wort}, \bibinfo{author}{R.~S. Balmer},
\newblock \bibinfo{title}{{Diamond as an electronic material}},
\newblock \bibinfo{journal}{Materials Today}  \bibinfo{volume}{11}
  (\bibinfo{year}{2008}) \bibinfo{pages}{22--28}.
  \DOIprefix\doi{10.1016/S1369-7021(07)70349-8}.
\bibitem[{Koike et~al.(1992)Koike, Parkin, and Mitchell}]{Koike1992}
\bibinfo{author}{J.~Koike}, \bibinfo{author}{D.~M. Parkin},
  \bibinfo{author}{T.~E. Mitchell},
\newblock \bibinfo{title}{{Displacement threshold energy for type IIa
  diamond}},
\newblock \bibinfo{journal}{Applied Physics Letters}  \bibinfo{volume}{60}
  (\bibinfo{year}{1992}) \bibinfo{pages}{1450--1452}.
  \DOIprefix\doi{10.1063/1.107267}.
\bibitem[{Zhang et~al.(2017)Zhang, Zhao, Weber, Nordlund, Granberg, and
  Djurabekova}]{ZHANG2017221}
\bibinfo{author}{Y.~Zhang}, \bibinfo{author}{S.~Zhao}, \bibinfo{author}{W.~J.
  Weber}, \bibinfo{author}{K.~Nordlund}, \bibinfo{author}{F.~Granberg},
  \bibinfo{author}{F.~Djurabekova},
\newblock \bibinfo{title}{Atomic-level heterogeneity and defect dynamics in
  concentrated solid-solution alloys},
\newblock \bibinfo{journal}{Current Opinion in Solid State and Materials
  Science}  \bibinfo{volume}{21} (\bibinfo{year}{2017})
  \bibinfo{pages}{221--237}.
  \DOIprefix\doi{https://doi.org/10.1016/j.cossms.2017.02.002},
  \bibinfo{note}{concentrated Solid Solution Alloys Perspective}.
\bibitem[{Zhang et~al.(2022)Zhang, Osetsky, and Weber}]{Zhang2022}
\bibinfo{author}{Y.~Zhang}, \bibinfo{author}{Y.~N. Osetsky},
  \bibinfo{author}{W.~J. Weber},
\newblock \bibinfo{title}{{Tunable Chemical Disorder in Concentrated Alloys:
  Defect Physics and Radiation Performance}},
\newblock \bibinfo{journal}{Chemical Reviews}  \bibinfo{volume}{122}
  (\bibinfo{year}{2022}) \bibinfo{pages}{789--829}.
  \DOIprefix\doi{10.1021/acs.chemrev.1c00387}.
\bibitem[{Zou et~al.(2020)Zou, Bohon, Smedley, Distel, Schmitt, Zhu, Zhang, and
  Muller}]{Zou2020}
\bibinfo{author}{M.~Zou}, \bibinfo{author}{J.~Bohon},
  \bibinfo{author}{J.~Smedley}, \bibinfo{author}{J.~Distel},
  \bibinfo{author}{K.~Schmitt}, \bibinfo{author}{R.~Y. Zhu}, et~al.,
\newblock \bibinfo{title}{{Proton radiation effects on carrier transport in
  diamond radiation detectors}},
\newblock \bibinfo{journal}{AIP Advances}  \bibinfo{volume}{10}
  (\bibinfo{year}{2020}) \bibinfo{pages}{1--7}.
  \DOIprefix\doi{10.1063/1.5130768}.
\bibitem[{{De Boer} et~al.(2007){De Boer}, Bol, Furgeri, M{\"{u}}ller, Sander,
  Berdermann, Pomorski, and Huhtinen}]{DeBoer2007}
\bibinfo{author}{W.~{De Boer}}, \bibinfo{author}{J.~Bol},
  \bibinfo{author}{A.~Furgeri}, \bibinfo{author}{S.~M{\"{u}}ller},
  \bibinfo{author}{C.~Sander}, \bibinfo{author}{E.~Berdermann}, et~al.,
\newblock \bibinfo{title}{{Radiation hardness of diamond and silicon sensors
  compared}},
\newblock \bibinfo{journal}{Physica Status Solidi (A) Applications and
  Materials Science}  \bibinfo{volume}{204} (\bibinfo{year}{2007})
  \bibinfo{pages}{3004--3010}. \DOIprefix\doi{10.1002/pssa.200776327}.
  \href{http://arxiv.org/abs/0705.0171}{{\tt arXiv:0705.0171}}.
\bibitem[{Liu et~al.(2020)Liu, Zhang, Wu, Qi, Hei, and Lv}]{Liu2020}
\bibinfo{author}{Y.~Liu}, \bibinfo{author}{J.~Zhang}, \bibinfo{author}{F.~Wu},
  \bibinfo{author}{M.~Qi}, \bibinfo{author}{L.~Hei}, \bibinfo{author}{F.~Lv},
\newblock \bibinfo{title}{{The Raman Spectroscopy and XPS investigation of CVD
  diamond after fast-neutron irradiation}},
\newblock \bibinfo{journal}{Materials Today Communications}
  \bibinfo{volume}{22} (\bibinfo{year}{2020}).
  \DOIprefix\doi{10.1016/j.mtcomm.2019.100699}.
\bibitem[{Nordlund et~al.(2018)Nordlund, Zinkle, Sand, Granberg, Averback,
  Stoller, Suzudo, Malerba, Banhart, Weber, Willaime, Dudarev, and
  Simeone}]{Nordlund2018}
\bibinfo{author}{K.~Nordlund}, \bibinfo{author}{S.~J. Zinkle},
  \bibinfo{author}{A.~E. Sand}, \bibinfo{author}{F.~Granberg},
  \bibinfo{author}{R.~S. Averback}, \bibinfo{author}{R.~E. Stoller}, et~al.,
\newblock \bibinfo{title}{{Primary radiation damage: A review of current
  understanding and models}},
\newblock \bibinfo{journal}{Journal of Nuclear Materials}
  \bibinfo{volume}{512} (\bibinfo{year}{2018}) \bibinfo{pages}{450--479}.
  \DOIprefix\doi{10.1016/j.jnucmat.2018.10.027}.
\bibitem[{Olivier et~al.(2018)Olivier, Neethling, Kroon, Naidoo, Allen, Sawada,
  {Van Aken}, and Kirkland}]{Olivier2018}
\bibinfo{author}{E.~J. Olivier}, \bibinfo{author}{J.~H. Neethling},
  \bibinfo{author}{R.~E. Kroon}, \bibinfo{author}{S.~R. Naidoo},
  \bibinfo{author}{C.~S. Allen}, \bibinfo{author}{H.~Sawada}, et~al.,
\newblock \bibinfo{title}{{Imaging the atomic structure and local chemistry of
  platelets in natural type Ia diamond}},
\newblock \bibinfo{journal}{Nature Materials}  \bibinfo{volume}{17}
  (\bibinfo{year}{2018}) \bibinfo{pages}{243--248}.
  \DOIprefix\doi{10.1038/s41563-018-0024-6}.
\bibitem[{N{\'{e}}meth et~al.(2020)N{\'{e}}meth, Mccoll, Garvie, Salzmann,
  Murri, and Mcmillan}]{Nemeth2020}
\bibinfo{author}{P.~N{\'{e}}meth}, \bibinfo{author}{K.~Mccoll},
  \bibinfo{author}{L.~A.~J. Garvie}, \bibinfo{author}{C.~G. Salzmann},
  \bibinfo{author}{M.~Murri}, \bibinfo{author}{P.~F. Mcmillan},
\newblock \bibinfo{title}{{Complex nanostructures in diamond}},
\newblock \bibinfo{journal}{Nature Materials}  \bibinfo{volume}{19}
  (\bibinfo{year}{2020}). \DOIprefix\doi{10.1038/s41563-020-0759-8}.
\bibitem[{N{\'{e}}meth et~al.(2014)N{\'{e}}meth, Garvie, Aoki, Dubrovinskaia,
  Dubrovinsky, and Buseck}]{Garvie2014}
\bibinfo{author}{P.~N{\'{e}}meth}, \bibinfo{author}{L.~A.~J. Garvie},
  \bibinfo{author}{T.~Aoki}, \bibinfo{author}{N.~Dubrovinskaia},
  \bibinfo{author}{L.~Dubrovinsky}, \bibinfo{author}{P.~R. Buseck},
\newblock \bibinfo{title}{{Lonsdaleite is faulted and twinned cubic diamond and
  does not exist as a discrete material}},
\newblock \bibinfo{journal}{Nature Communications}  \bibinfo{volume}{5}
  (\bibinfo{year}{2014}) \bibinfo{pages}{5447}.
  \DOIprefix\doi{10.1038/ncomms6447}.
\bibitem[{Nshingabigwi et~al.(2014)Nshingabigwi, Derry, Naidoo, Neethling,
  Olivier, O'Connell, and Levitt}]{nshingabigwi_electron_2014}
\bibinfo{author}{E.~Nshingabigwi}, \bibinfo{author}{T.~Derry},
  \bibinfo{author}{S.~Naidoo}, \bibinfo{author}{J.~Neethling},
  \bibinfo{author}{E.~Olivier}, \bibinfo{author}{J.~O'Connell}, et~al.,
\newblock \bibinfo{title}{Electron microscopy profiling of ion implantation
  damage in diamond: {Dependence} on fluence and annealing},
\newblock \bibinfo{journal}{Diamond and Related Materials}
  \bibinfo{volume}{49} (\bibinfo{year}{2014}) \bibinfo{pages}{1--8}.
  \DOIprefix\doi{10.1016/j.diamond.2014.07.010}.
\bibitem[{Bulavin and Kulikov(2018)}]{Bulavin2018}
\bibinfo{author}{M.~Bulavin}, \bibinfo{author}{S.~Kulikov},
\newblock \bibinfo{title}{{Current experiments at the irradiation facility of
  the IBR-2 reactor}},
\newblock \bibinfo{journal}{Journal of Physics: Conference Series}
  \bibinfo{volume}{1021} (\bibinfo{year}{2018}).
  \DOIprefix\doi{10.1088/1742-6596/1021/1/012041}.
\bibitem[{Rutherfoord(2012)}]{Rutherfoord2012}
\bibinfo{author}{J.~Rutherfoord},
\newblock \bibinfo{title}{{Upgrade plans for the ATLAS Forward Calorimeter at
  the HL-LHC}},
\newblock \bibinfo{journal}{Journal of Physics: Conference Series}
  \bibinfo{volume}{404} (\bibinfo{year}{2012}) \bibinfo{pages}{012015}.
  \DOIprefix\doi{10.1088/1742-6596/404/1/012015}.
\bibitem[{García~Alía et~al.(2018)García~Alía, Brugger, Cerutti, Danzeca,
  Ferrari, Gilardoni, Kadi, Kastriotou, Lechner, Martinella, Stein, Thurel,
  Tsinganis, and Uznanski}]{8116686}
\bibinfo{author}{R.~García~Alía}, \bibinfo{author}{M.~Brugger},
  \bibinfo{author}{F.~Cerutti}, \bibinfo{author}{S.~Danzeca},
  \bibinfo{author}{A.~Ferrari}, \bibinfo{author}{S.~Gilardoni}, et~al.,
\newblock \bibinfo{title}{Lhc and hl-lhc: Present and future radiation
  environment in the high-luminosity collision points and rha implications},
\newblock \bibinfo{journal}{IEEE Transactions on Nuclear Science}
  \bibinfo{volume}{65} (\bibinfo{year}{2018}) \bibinfo{pages}{448--456}.
  \DOIprefix\doi{10.1109/TNS.2017.2776107}.
\bibitem[{B{\'{e}}jar et~al.(2020)B{\'{e}}jar, Br{\"{u}}ning, Fessia, Lamont,
  Rossi, Tavian, and Zerlauth}]{Bejar2020}
\bibinfo{author}{I.~B{\'{e}}jar}, \bibinfo{author}{A.~O. Br{\"{u}}ning},
  \bibinfo{author}{P.~Fessia}, \bibinfo{author}{M.~Lamont},
  \bibinfo{author}{L.~Rossi}, \bibinfo{author}{L.~Tavian}, et~al.,
  \bibinfo{title}{{High-Luminosity Large Hadron Collider (HL-LHC) Technical
  design report}},  \bibinfo{year}{2020}.
  \DOIprefix\doi{https://doi.org/10.23731/CYRM-2020-0010}.
\bibitem[{Hei et~al.(2012)Hei, Liu, Li, Song, Tang, and Lu}]{Hei2012}
\bibinfo{author}{L.~F. Hei}, \bibinfo{author}{J.~Liu}, \bibinfo{author}{C.~M.
  Li}, \bibinfo{author}{J.~H. Song}, \bibinfo{author}{W.~Z. Tang},
  \bibinfo{author}{F.~X. Lu},
\newblock \bibinfo{title}{{Fabrication and characterizations of large
  homoepitaxial single crystal diamond grown by DC arc plasma jet CVD}},
\newblock \bibinfo{journal}{Diamond and Related Materials}
  \bibinfo{volume}{30} (\bibinfo{year}{2012}) \bibinfo{pages}{77--84}.
  \DOIprefix\doi{10.1016/j.diamond.2012.10.002}.
\bibitem[{Bulavin et~al.(2015)Bulavin, Cheplakov, Kukhtin, Kulagin, Kulikov,
  Shabalin, and Verkhoglyadov}]{bulavin_irradiation_2015}
\bibinfo{author}{M.~Bulavin}, \bibinfo{author}{A.~Cheplakov},
  \bibinfo{author}{V.~Kukhtin}, \bibinfo{author}{E.~Kulagin},
  \bibinfo{author}{S.~Kulikov}, \bibinfo{author}{E.~Shabalin}, et~al.,
\newblock \bibinfo{title}{Irradiation facility at the {IBR}-2 reactor for
  investigation of material radiation hardness},
\newblock \bibinfo{journal}{Nuclear Instruments and Methods in Physics Research
  Section B: Beam Interactions with Materials and Atoms}  \bibinfo{volume}{343}
  (\bibinfo{year}{2015}) \bibinfo{pages}{26--29}.
  \DOIprefix\doi{10.1016/j.nimb.2014.11.024}.
\bibitem[{Hickey et~al.(2006)Hickey, Kuryliw, Siebein, Jones, Chodelka, and
  Elliman}]{hickey_cross-sectional_2006}
\bibinfo{author}{D.~P. Hickey}, \bibinfo{author}{E.~Kuryliw},
  \bibinfo{author}{K.~Siebein}, \bibinfo{author}{K.~S. Jones},
  \bibinfo{author}{R.~Chodelka}, \bibinfo{author}{R.~Elliman},
\newblock \bibinfo{title}{Cross-sectional transmission electron microscopy
  method and studies of implant damage in single crystal diamond},
\newblock \bibinfo{journal}{Journal of Vacuum Science \& Technology A: Vacuum,
  Surfaces, and Films}  \bibinfo{volume}{24} (\bibinfo{year}{2006})
  \bibinfo{pages}{1302--1307}. \DOIprefix\doi{10.1116/1.2209659}.
\bibitem[{Rubanov and Suvorova(2011)}]{rubanov_study_2011}
\bibinfo{author}{S.~Rubanov}, \bibinfo{author}{A.~Suvorova},
\newblock \bibinfo{title}{The {Study} of the {FIB} {Induced} {Damage} in
  {Diamond}},
\newblock \bibinfo{journal}{Microscopy and Microanalysis}  \bibinfo{volume}{17}
  (\bibinfo{year}{2011}) \bibinfo{pages}{700--701}.
  \DOIprefix\doi{10.1017/S1431927611004375}.
\bibitem[{Rubanov(2016)}]{fib2}
\bibinfo{author}{S.~Rubanov}, \bibinfo{title}{Optimisation of the FIB induced
  damage in TEM diamond samples}, \bibinfo{year}{2016},  pp.
  \bibinfo{pages}{358--359}.
  \DOIprefix\doi{10.1002/9783527808465.EMC2016.6696}.
\bibitem[{Schneider et~al.(2012)Schneider, Rasband, and
  Eliceiri}]{schneider_nih_2012}
\bibinfo{author}{C.~A. Schneider}, \bibinfo{author}{W.~S. Rasband},
  \bibinfo{author}{K.~W. Eliceiri},
\newblock \bibinfo{title}{{NIH} {Image} to {ImageJ}: 25 years of image
  analysis},
\newblock \bibinfo{journal}{Nature Methods}  \bibinfo{volume}{9}
  (\bibinfo{year}{2012}) \bibinfo{pages}{671--675}.
  \DOIprefix\doi{10.1038/nmeth.2089}, \bibinfo{note}{publisher: Nature
  Publishing Group}.
\bibitem[{H{\"{y}}tch et~al.(1998)H{\"{y}}tch, Snoeck, and Kilaas}]{Hytch1998}
\bibinfo{author}{M.~J. H{\"{y}}tch}, \bibinfo{author}{E.~Snoeck},
  \bibinfo{author}{R.~Kilaas},
\newblock \bibinfo{title}{{Quantitative measurement of displacement and strain
  fields from HREM micrographs}},
\newblock \bibinfo{journal}{Ultramicroscopy}  \bibinfo{volume}{74}
  (\bibinfo{year}{1998}) \bibinfo{pages}{131--146}.
  \DOIprefix\doi{10.1016/S0304-3991(98)00035-7}.
\bibitem[{Agostinelli et~al.(2003)Agostinelli, Allison, Amako, Apostolakis,
  Araujo, Arce, Asai, Axen, Banerjee, Barrand, Behner, Bellagamba, Boudreau,
  Broglia, Brunengo, Burkhardt, Chauvie, Chuma, Chytracek, Cooperman, Cosmo,
  Degtyarenko, Dell'Acqua, Depaola, Dietrich, Enami, Feliciello, Ferguson,
  Fesefeldt, Folger, Foppiano, Forti, Garelli, Giani, Giannitrapani, Gibin,
  {Gomez Cadenas}, Gonzalez, {Gracia Abril}, Greeniaus, Greiner, Grichine,
  Grossheim, Guatelli, Gumplinger, Hamatsu, Hashimoto, Hasui, Heikkinen,
  Howard, Ivanchenko, Johnson, Jones, Kallenbach, Kanaya, Kawabata, Kawabata,
  Kawaguti, Kelner, Kent, Kimura, Kodama, Kokoulin, Kossov, Kurashige, Lamanna,
  Lampen, Lara, Lefebure, Lei, Liendl, Lockman, Longo, Magni, Maire, Medernach,
  Minamimoto, {Mora de Freitas}, Morita, Murakami, Nagamatu, Nartallo,
  Nieminen, Nishimura, Ohtsubo, Okamura, O'Neale, Oohata, Paech, Perl,
  Pfeiffer, Pia, Ranjard, Rybin, Sadilov, di~Salvo, Santin, Sasaki, Savvas,
  Sawada, Scherer, Sei, Sirotenko, Smith, Starkov, Stoecker, Sulkimo, Takahata,
  Tanaka, Tcherniaev, {Safai Tehrani}, Tropeano, Truscott, Uno, Urban, Urban,
  Verderi, Walkden, Wander, Weber, Wellisch, Wenaus, Williams, Wright, Yamada,
  Yoshida, and Zschiesche}]{Agostinelli2003}
\bibinfo{author}{S.~Agostinelli}, \bibinfo{author}{J.~Allison},
  \bibinfo{author}{K.~Amako}, \bibinfo{author}{J.~Apostolakis},
  \bibinfo{author}{H.~Araujo}, \bibinfo{author}{P.~Arce}, et~al.,
\newblock \bibinfo{title}{{GEANT4 - A simulation toolkit}},
\newblock \bibinfo{journal}{Nuclear Instruments and Methods in Physics
  Research, Section A: Accelerators, Spectrometers, Detectors and Associated
  Equipment}  \bibinfo{volume}{506} (\bibinfo{year}{2003})
  \bibinfo{pages}{250--303}. \DOIprefix\doi{10.1016/S0168-9002(03)01368-8}.
\bibitem[{Born and Mayer(1932)}]{Born1932}
\bibinfo{author}{M.~Born}, \bibinfo{author}{J.~E. Mayer},
\newblock \bibinfo{title}{Zur gittertheorie der ionenkristalle},
\newblock \bibinfo{journal}{Zeitschrift für Physik}  \bibinfo{volume}{75}
  (\bibinfo{year}{1932}) \bibinfo{pages}{1--18}.
  \DOIprefix\doi{10.1007/BF01340511}.
\bibitem[{Was(2017)}]{was2016fundamentals}
\bibinfo{author}{G.~S. Was},
\newblock \bibinfo{title}{{Dislocation Microstructure}},
\newblock in: \bibinfo{booktitle}{Fundamentals of Radiation Materials Science},
  \bibinfo{publisher}{Springer New York}, \bibinfo{address}{New York, NY},
  \bibinfo{year}{2017},  pp. \bibinfo{pages}{301--378}.
  \DOIprefix\doi{10.1007/978-1-4939-3438-6_7}.
\bibitem[{Thompson et~al.(2022)Thompson, Aktulga, Berger, Bolintineanu, Brown,
  Crozier, in~'t Veld, Kohlmeyer, Moore, Nguyen, Shan, Stevens, Tranchida,
  Trott, and Plimpton}]{LAMMPS}
\bibinfo{author}{A.~P. Thompson}, \bibinfo{author}{H.~M. Aktulga},
  \bibinfo{author}{R.~Berger}, \bibinfo{author}{D.~S. Bolintineanu},
  \bibinfo{author}{W.~M. Brown}, \bibinfo{author}{P.~S. Crozier}, et~al.,
\newblock \bibinfo{title}{{LAMMPS} - a flexible simulation tool for
  particle-based materials modeling at the atomic, meso, and continuum scales},
\newblock \bibinfo{journal}{Comp. Phys. Comm.}  \bibinfo{volume}{271}
  (\bibinfo{year}{2022}) \bibinfo{pages}{108171}.
  \DOIprefix\doi{10.1016/j.cpc.2021.108171}.
\bibitem[{Tersoff(1988)}]{Tersoff1988}
\bibinfo{author}{J.~Tersoff},
\newblock \bibinfo{title}{{New empirical approach for the structure and energy
  of covalent systems}},
\newblock \bibinfo{journal}{Physical Review B}  \bibinfo{volume}{37}
  (\bibinfo{year}{1988}) \bibinfo{pages}{6991--7000}.
  \DOIprefix\doi{10.1103/PhysRevB.37.6991}.
\bibitem[{Ziegler and Biersack(1985)}]{Ziegler1985}
\bibinfo{author}{J.~F. Ziegler}, \bibinfo{author}{J.~P. Biersack},
  \bibinfo{title}{The Stopping and Range of Ions in Matter},
  \bibinfo{publisher}{Springer US}, \bibinfo{address}{Boston, MA},
  \bibinfo{year}{1985},  pp. \bibinfo{pages}{93--129}.
  \DOIprefix\doi{10.1007/978-1-4615-8103-1_3}.
\bibitem[{Madsen and Susi(2021)}]{madsen_abtem_2021}
\bibinfo{author}{J.~Madsen}, \bibinfo{author}{T.~Susi},
\newblock \bibinfo{title}{{{abTEM}}: Transmission electron microscopy from
  first principles},
\newblock \bibinfo{journal}{Open Research Europe}  \bibinfo{volume}{1}
  (\bibinfo{year}{2021}) \bibinfo{pages}{13015}.
  \DOIprefix\doi{10.12688/openreseurope.13015.1}.
\bibitem[{Peters(2021)}]{PETERS2021113364}
\bibinfo{author}{J.~J. Peters},
\newblock \bibinfo{title}{A fast frozen phonon algorithm using mixed static
  potentials},
\newblock \bibinfo{journal}{Ultramicroscopy}  \bibinfo{volume}{229}
  (\bibinfo{year}{2021}) \bibinfo{pages}{113364}.
  \DOIprefix\doi{https://doi.org/10.1016/j.ultramic.2021.113364}.
\bibitem[{Poklonski et~al.(2023)Poklonski, Khomich, Svito, Vyrko, Poklonskaya,
  Kovalev, Kozlova, Khmelnitskii, and Khomich}]{poklonski_magnetic_2023}
\bibinfo{author}{N.~A. Poklonski}, \bibinfo{author}{A.~A. Khomich},
  \bibinfo{author}{I.~A. Svito}, \bibinfo{author}{S.~A. Vyrko},
  \bibinfo{author}{O.~N. Poklonskaya}, \bibinfo{author}{A.~I. Kovalev}, et~al.,
\newblock \bibinfo{title}{Magnetic and {Optical} {Properties} of {Natural}
  {Diamonds} with {Subcritical} {Radiation} {Damage} {Induced} by {Fast}
  {Neutrons}},
\newblock \bibinfo{journal}{Applied Sciences}  \bibinfo{volume}{13}
  (\bibinfo{year}{2023}) \bibinfo{pages}{6221}.
  \DOIprefix\doi{10.3390/app13106221}.
\bibitem[{Khomich et~al.(2020)Khomich, Khmelnitsky, and
  Khomich}]{khomich_probing_2020}
\bibinfo{author}{A.~A. Khomich}, \bibinfo{author}{R.~A. Khmelnitsky},
  \bibinfo{author}{A.~V. Khomich},
\newblock \bibinfo{title}{Probing the {Nanostructure} of {Neutron}-{Irradiated}
  {Diamond} {Using} {Raman} {Spectroscopy}},
\newblock \bibinfo{journal}{Nanomaterials}  \bibinfo{volume}{10}
  (\bibinfo{year}{2020}) \bibinfo{pages}{1166}.
  \DOIprefix\doi{10.3390/nano10061166}.
\bibitem[{Poklonskaya et~al.(2015)Poklonskaya, Vyrko, Khomich, Averin, Khomich,
  Khmelnitsky, and Poklonskia}]{poklonskaya_raman_2015}
\bibinfo{author}{O.~N. Poklonskaya}, \bibinfo{author}{S.~A. Vyrko},
  \bibinfo{author}{A.~A. Khomich}, \bibinfo{author}{A.~A. Averin},
  \bibinfo{author}{A.~V. Khomich}, \bibinfo{author}{R.~A. Khmelnitsky}, et~al.,
\newblock \bibinfo{title}{Raman {Scattering} in {Natural} {Diamond} {Crystals}
  {Implanted} with {High}-{Energy} {Ions} and {Irradiated} with {Fast}
  {Neutrons}},
\newblock \bibinfo{journal}{Journal of Applied Spectroscopy}
  \bibinfo{volume}{81} (\bibinfo{year}{2015}) \bibinfo{pages}{969--977}.
  \DOIprefix\doi{10.1007/s10812-015-0037-8}.
\bibitem[{Poklonskaya and Khomich(2013)}]{poklonskaya_raman_2013}
\bibinfo{author}{O.~N. Poklonskaya}, \bibinfo{author}{A.~A. Khomich},
\newblock \bibinfo{title}{Raman {Scattering} in a {Diamond} {Crystal}
  {Implanted} by {High}-{Energy} {Nickel} {Ions}},
\newblock \bibinfo{journal}{Journal of Applied Spectroscopy}
  \bibinfo{volume}{80} (\bibinfo{year}{2013}) \bibinfo{pages}{715--720}.
  \DOIprefix\doi{10.1007/s10812-013-9831-3}.
\bibitem[{Leech et~al.(2004)Leech, Reeves, Holland, and
  Ridgway}]{leech_effect_2004}
\bibinfo{author}{P.~W. Leech}, \bibinfo{author}{G.~K. Reeves},
  \bibinfo{author}{A.~Holland}, \bibinfo{author}{M.~C. Ridgway},
\newblock \bibinfo{title}{The effect of {Au} and {O} implantation on the etch
  rate of {CVD} diamond},
\newblock \bibinfo{journal}{Applied Surface Science}  \bibinfo{volume}{221}
  (\bibinfo{year}{2004}) \bibinfo{pages}{302--307}.
  \DOIprefix\doi{10.1016/S0169-4332(03)00949-8}.
\bibitem[{Goss et~al.(2001)Goss, Coomer, Jones, Shaw, Briddon, Rayson, and
  Öberg}]{goss_self-interstitial_2001}
\bibinfo{author}{J.~P. Goss}, \bibinfo{author}{B.~J. Coomer},
  \bibinfo{author}{R.~Jones}, \bibinfo{author}{T.~D. Shaw},
  \bibinfo{author}{P.~R. Briddon}, \bibinfo{author}{M.~Rayson}, et~al.,
\newblock \bibinfo{title}{Self-interstitial aggregation in diamond},
\newblock \bibinfo{journal}{Physical Review B}  \bibinfo{volume}{63}
  (\bibinfo{year}{2001}) \bibinfo{pages}{195208}.
  \DOIprefix\doi{10.1103/PhysRevB.63.195208}.
\bibitem[{Kalish et~al.(1999)Kalish, Reznik, Nugent, and
  Prawer}]{kalish_nature_1999}
\bibinfo{author}{R.~Kalish}, \bibinfo{author}{A.~Reznik},
  \bibinfo{author}{K.~Nugent}, \bibinfo{author}{S.~Prawer},
\newblock \bibinfo{title}{The nature of damage in ion-implanted and annealed
  diamond},
\newblock \bibinfo{journal}{Nuclear Instruments and Methods in Physics Research
  Section B: Beam Interactions with Materials and Atoms}  \bibinfo{volume}{148}
  (\bibinfo{year}{1999}) \bibinfo{pages}{626--633}.
  \DOIprefix\doi{10.1016/S0168-583X(98)00857-X}.
\bibitem[{Prawer et~al.(2000)Prawer, Nugent, Jamieson, Orwa, Bursill, and
  Peng}]{prawer_raman_2000}
\bibinfo{author}{S.~Prawer}, \bibinfo{author}{K.~Nugent},
  \bibinfo{author}{D.~Jamieson}, \bibinfo{author}{J.~Orwa},
  \bibinfo{author}{L.~Bursill}, \bibinfo{author}{J.~Peng},
\newblock \bibinfo{title}{The {Raman} spectrum of nanocrystalline diamond},
\newblock \bibinfo{journal}{Chemical Physics Letters}  \bibinfo{volume}{332}
  (\bibinfo{year}{2000}) \bibinfo{pages}{93--97}.
  \DOIprefix\doi{10.1016/S0009-2614(00)01236-7}.
\bibitem[{Orwa et~al.(2000)Orwa, Nugent, Jamieson, and
  Prawer}]{orwa_raman_2000}
\bibinfo{author}{J.~O. Orwa}, \bibinfo{author}{K.~W. Nugent},
  \bibinfo{author}{D.~N. Jamieson}, \bibinfo{author}{S.~Prawer},
\newblock \bibinfo{title}{Raman investigation of damage caused by deep ion
  implantation in diamond},
\newblock \bibinfo{journal}{Physical Review B}  \bibinfo{volume}{62}
  (\bibinfo{year}{2000}) \bibinfo{pages}{5461--5472}.
  \DOIprefix\doi{10.1103/PhysRevB.62.5461}.
\bibitem[{Ferrari and Robertson(2000)}]{ferrari_interpretation_2000}
\bibinfo{author}{A.~C. Ferrari}, \bibinfo{author}{J.~Robertson},
\newblock \bibinfo{title}{Interpretation of {Raman} spectra of disordered and
  amorphous carbon},
\newblock \bibinfo{journal}{Physical Review B}  \bibinfo{volume}{61}
  (\bibinfo{year}{2000}) \bibinfo{pages}{14095--14107}.
  \DOIprefix\doi{10.1103/PhysRevB.61.14095}.
\bibitem[{Khomich et~al.(2019)Khomich, Averin, Poklonskaya, Bokova-Sirosh,
  Dzeraviaha, Khmelnitsky, Vlasov, Shenderova, Poklonski, and
  Khomich}]{khomich_features_2019}
\bibinfo{author}{A.~A. Khomich}, \bibinfo{author}{A.~A. Averin},
  \bibinfo{author}{O.~N. Poklonskaya}, \bibinfo{author}{S.~N. Bokova-Sirosh},
  \bibinfo{author}{A.~N. Dzeraviaha}, \bibinfo{author}{R.~A. Khmelnitsky},
  et~al.,
\newblock \bibinfo{title}{Features of the 1640 $cm^{-1}$ band in the raman
  spectra of radiation-damaged and nano-sized diamonds},
\newblock \bibinfo{journal}{Journal of Physics: Conference Series}
  \bibinfo{volume}{1400} (\bibinfo{year}{2019}) \bibinfo{pages}{044017}.
  \DOIprefix\doi{10.1088/1742-6596/1400/4/044017}.
\bibitem[{Muller et~al.(1993)Muller, Tzou, Raj, and Silcox}]{Muller1993}
\bibinfo{author}{D.~A. Muller}, \bibinfo{author}{Y.~Tzou},
  \bibinfo{author}{R.~Raj}, \bibinfo{author}{J.~Silcox},
\newblock \bibinfo{title}{{Mapping sp2 and sp3 states of carbon at
  sub-nanometre spatial resolution}},
\newblock \bibinfo{journal}{Nature}  \bibinfo{volume}{366}
  (\bibinfo{year}{1993}) \bibinfo{pages}{725--727}.
  \DOIprefix\doi{10.1038/366725a0}.
\bibitem[{Rani et~al.(2018)Rani, Panda, Kumar, and Titovich}]{Rani2018}
\bibinfo{author}{R.~Rani}, \bibinfo{author}{K.~Panda},
  \bibinfo{author}{N.~Kumar}, \bibinfo{author}{K.~A. Titovich},
\newblock \bibinfo{title}{{Tribological Properties of Ultrananocrystalline
  Diamond Films : Mechanochemical Transformation of Sliding Interfaces}},
\newblock \bibinfo{journal}{Scientific Reports}   (\bibinfo{year}{2018})
  \bibinfo{pages}{1--16}. \DOIprefix\doi{10.1038/s41598-017-18425-4}.
\bibitem[{Bruley et~al.(1995)Bruley, Williams, Cuomo, and
  Pappas}]{bruley_quantitative_1995}
\bibinfo{author}{J.~Bruley}, \bibinfo{author}{D.~B. Williams},
  \bibinfo{author}{J.~J. Cuomo}, \bibinfo{author}{D.~P. Pappas},
\newblock \bibinfo{title}{Quantitative near‐edge structure analysis of
  diamond‐like carbon in the electron microscope using a two‐window
  method},
\newblock \bibinfo{journal}{Journal of Microscopy}  \bibinfo{volume}{180}
  (\bibinfo{year}{1995}) \bibinfo{pages}{22--32}.
  \DOIprefix\doi{10.1111/j.1365-2818.1995.tb03653.x}.
\bibitem[{Blumenau et~al.(2003)Blumenau, Jones, Frauenheim, Willems, Lebedev,
  {Van Tendeloo}, Fisher, and Martineau}]{Blumenau2003}
\bibinfo{author}{T.~Blumenau}, \bibinfo{author}{R.~Jones},
  \bibinfo{author}{T.~Frauenheim}, \bibinfo{author}{B.~Willems},
  \bibinfo{author}{I.~Lebedev}, \bibinfo{author}{G.~{Van Tendeloo}}, et~al.,
\newblock \bibinfo{title}{{Dislocations in diamond: Dissociation into partials
  and their glide motion}},
\newblock \bibinfo{journal}{Physical Review B - Condensed Matter and Materials
  Physics}  \bibinfo{volume}{68} (\bibinfo{year}{2003}) \bibinfo{pages}{1--9}.
  \DOIprefix\doi{10.1103/PhysRevB.68.014115}.
\bibitem[{Lomer(1951)}]{Lomer1951}
\bibinfo{author}{W.~Lomer},
\newblock \bibinfo{title}{{A dislocation reaction in the face-centred cubic
  lattice}},
\newblock \bibinfo{journal}{The London, Edinburgh, and Dublin Philosophical
  Magazine and Journal of Science}  \bibinfo{volume}{42} (\bibinfo{year}{1951})
  \bibinfo{pages}{1327--1331}. \DOIprefix\doi{10.1080/14786444108561389}.
\bibitem[{Cottrell(1952)}]{Cottrell1952}
\bibinfo{author}{A.~Cottrell},
\newblock \bibinfo{title}{{The formation of immobile dislocations during
  slip}},
\newblock \bibinfo{journal}{The London, Edinburgh, and Dublin Philosophical
  Magazine and Journal of Science}  \bibinfo{volume}{43} (\bibinfo{year}{1952})
  \bibinfo{pages}{645--647}. \DOIprefix\doi{10.1080/14786440608520220}.
\bibitem[{Abu-Odeh et~al.(2022)Abu-Odeh, Allaparti, and Asta}]{Abu-Odeh2022}
\bibinfo{author}{A.~Abu-Odeh}, \bibinfo{author}{T.~Allaparti},
  \bibinfo{author}{M.~Asta},
\newblock \bibinfo{title}{{Structure and glide of Lomer and Lomer-Cottrell
  dislocations: Atomistic simulations for model concentrated alloy solid
  solutions}},
\newblock \bibinfo{journal}{Physical Review Materials}  \bibinfo{volume}{6}
  (\bibinfo{year}{2022}) \bibinfo{pages}{1--12}.
  \DOIprefix\doi{10.1103/PhysRevMaterials.6.103603}.
\bibitem[{Zhang et~al.(2018)Zhang, Hattar, Chen, Shao, Li, Sun, Yu, Li, Taheri,
  Wang, Wang, and Nastasi}]{Zhang2018}
\bibinfo{author}{X.~Zhang}, \bibinfo{author}{K.~Hattar},
  \bibinfo{author}{Y.~Chen}, \bibinfo{author}{L.~Shao},
  \bibinfo{author}{J.~Li}, \bibinfo{author}{C.~Sun}, et~al.,
\newblock \bibinfo{title}{{Radiation damage in nanostructured materials}},
\newblock \bibinfo{journal}{Progress in Materials Science}
  \bibinfo{volume}{96} (\bibinfo{year}{2018}) \bibinfo{pages}{217--321}.
  \DOIprefix\doi{10.1016/j.pmatsci.2018.03.002}.
\bibitem[{Saada et~al.(1998)Saada, Adler, and Kalish}]{Saada1998}
\bibinfo{author}{D.~Saada}, \bibinfo{author}{J.~Adler},
  \bibinfo{author}{R.~Kalish},
\newblock \bibinfo{title}{{Transformation of diamond (sp3) to graphite (sp2)
  bonds by ion-impact}},
\newblock \bibinfo{journal}{International Journal of Modern Physics C}
  \bibinfo{volume}{9} (\bibinfo{year}{1998}) \bibinfo{pages}{61--69}.
  \DOIprefix\doi{10.1142/S0129183198000066}.
\bibitem[{Breuer and Briddon(1995)}]{Breuer1995}
\bibinfo{author}{S.~J. Breuer}, \bibinfo{author}{P.~R. Briddon},
\newblock \bibinfo{title}{{Ab initio investigation of the native defects in
  diamond and self-diffusion}},
\newblock \bibinfo{journal}{Physical Review B}  \bibinfo{volume}{51}
  (\bibinfo{year}{1995}) \bibinfo{pages}{6984--6994}.
  \DOIprefix\doi{10.1103/PhysRevB.51.6984}.
\bibitem[{{De Backer} et~al.(2016){De Backer}, van~den Bos, {Van den Broek},
  Sijbers, and {Van Aert}}]{DeBacker2016}
\bibinfo{author}{A.~{De Backer}}, \bibinfo{author}{K.~H. van~den Bos},
  \bibinfo{author}{W.~{Van den Broek}}, \bibinfo{author}{J.~Sijbers},
  \bibinfo{author}{S.~{Van Aert}},
\newblock \bibinfo{title}{{StatSTEM: An efficient approach for accurate and
  precise model-based quantification of atomic resolution electron microscopy
  images}},
\newblock \bibinfo{journal}{Ultramicroscopy}  \bibinfo{volume}{171}
  (\bibinfo{year}{2016}) \bibinfo{pages}{104--116}.
  \DOIprefix\doi{10.1016/j.ultramic.2016.08.018}.
\bibitem[{{Van Aert} et~al.(2009){Van Aert}, Verbeeck, Erni, Bals, Luysberg,
  Dyck, and Tendeloo}]{VanAert2009}
\bibinfo{author}{S.~{Van Aert}}, \bibinfo{author}{J.~Verbeeck},
  \bibinfo{author}{R.~Erni}, \bibinfo{author}{S.~Bals},
  \bibinfo{author}{M.~Luysberg}, \bibinfo{author}{D.~V. Dyck}, et~al.,
\newblock \bibinfo{title}{{Quantitative atomic resolution mapping using
  high-angle annular dark field scanning transmission electron microscopy}},
\newblock \bibinfo{journal}{Ultramicroscopy}  \bibinfo{volume}{109}
  (\bibinfo{year}{2009}) \bibinfo{pages}{1236--1244}.
  \DOIprefix\doi{10.1016/j.ultramic.2009.05.010}.
\bibitem[{Pennycook and Boatner(1988)}]{Pennycook1988}
\bibinfo{author}{S.~J. Pennycook}, \bibinfo{author}{L.~A. Boatner},
\newblock \bibinfo{title}{{Chemically sensitive structure-imaging with a
  scanning transmission electron microscope}},
\newblock \bibinfo{journal}{Nature}  \bibinfo{volume}{336}
  (\bibinfo{year}{1988}) \bibinfo{pages}{565--567}.
  \DOIprefix\doi{10.1038/336565a0}.
\bibitem[{Pennycook and Jesson(1990)}]{Pennycook1990}
\bibinfo{author}{S.~Pennycook}, \bibinfo{author}{D.~Jesson},
\newblock \bibinfo{title}{{High-resolution incoherent imaging of crystals}},
\newblock \bibinfo{journal}{Physical Review Letters}  \bibinfo{volume}{64}
  (\bibinfo{year}{1990}) \bibinfo{pages}{938--941}.
  \DOIprefix\doi{10.1103/PhysRevLett.64.938}.
\bibitem[{Pennycook and Jesson(1991)}]{Pennycook1991}
\bibinfo{author}{S.~Pennycook}, \bibinfo{author}{D.~Jesson},
\newblock \bibinfo{title}{{High-resolution Z-contrast imaging of crystals}},
\newblock \bibinfo{journal}{Ultramicroscopy}  \bibinfo{volume}{37}
  (\bibinfo{year}{1991}) \bibinfo{pages}{14--38}.
  \DOIprefix\doi{10.1016/0304-3991(91)90004-P}.
\bibitem[{Robinson and Torrens(1974)}]{Robinson1974}
\bibinfo{author}{M.~Robinson}, \bibinfo{author}{I.~Torrens},
\newblock \bibinfo{title}{{Computer simulation of atomic-displacement cascades
  in solids in the binary-collision approximation}},
\newblock \bibinfo{journal}{Physical Review B}  \bibinfo{volume}{9}
  (\bibinfo{year}{1974}) \bibinfo{pages}{5008--5024}.
  \DOIprefix\doi{10.1103/PhysRevB.9.5008}.
\bibitem[{Nordlund et~al.(2018)Nordlund, Zinkle, Sand, Granberg, Averback,
  Stoller, Suzudo, Malerba, Banhart, Weber, Willaime, Dudarev, and
  Simeone}]{Nordlund2018NC}
\bibinfo{author}{K.~Nordlund}, \bibinfo{author}{S.~J. Zinkle},
  \bibinfo{author}{A.~E. Sand}, \bibinfo{author}{F.~Granberg},
  \bibinfo{author}{R.~S. Averback}, \bibinfo{author}{R.~Stoller}, et~al.,
\newblock \bibinfo{title}{{Improving atomic displacement and replacement
  calculations with physically realistic damage models}},
\newblock \bibinfo{journal}{Nature Communications}  \bibinfo{volume}{9}
  (\bibinfo{year}{2018}) \bibinfo{pages}{1084}.
  \DOIprefix\doi{10.1038/s41467-018-03415-5}.
\bibitem[{Walker(1977)}]{walker_optical_1977}
\bibinfo{author}{J.~Walker},
\newblock \bibinfo{title}{An optical study of the {TR12} and {3H} defects in
  irradiated diamond},
\newblock \bibinfo{journal}{Journal of Physics C: Solid State Physics}
  \bibinfo{volume}{10} (\bibinfo{year}{1977}) \bibinfo{pages}{3031--3037}.
  \DOIprefix\doi{10.1088/0022-3719/10/16/013}.
\bibitem[{Hunt et~al.(2000)Hunt, Twitchen, Newton, Baker, Anthony, Banholzer,
  and Vagarali}]{prb67}
\bibinfo{author}{D.~C. Hunt}, \bibinfo{author}{D.~J. Twitchen},
  \bibinfo{author}{M.~E. Newton}, \bibinfo{author}{J.~M. Baker},
  \bibinfo{author}{T.~R. Anthony}, \bibinfo{author}{W.~F. Banholzer}, et~al.,
\newblock \bibinfo{title}{Identification of the neutral carbon 〈100〉-split
  interstitial in diamond},
\newblock \bibinfo{journal}{Phys. Rev. B}  \bibinfo{volume}{61}
  (\bibinfo{year}{2000}) \bibinfo{pages}{3863--3876}.
  \DOIprefix\doi{10.1103/PhysRevB.61.3863}.

\end{thebibliography}

\newpage
\appendix
\setcounter{appendixfigure}{0} 
\renewcommand{\thefigure}{\theappendixfigure} 
\section*{Supplementary Data}
\begin{figure}[H]
\label{figS1}
\centering
\includegraphics[width=\linewidth]{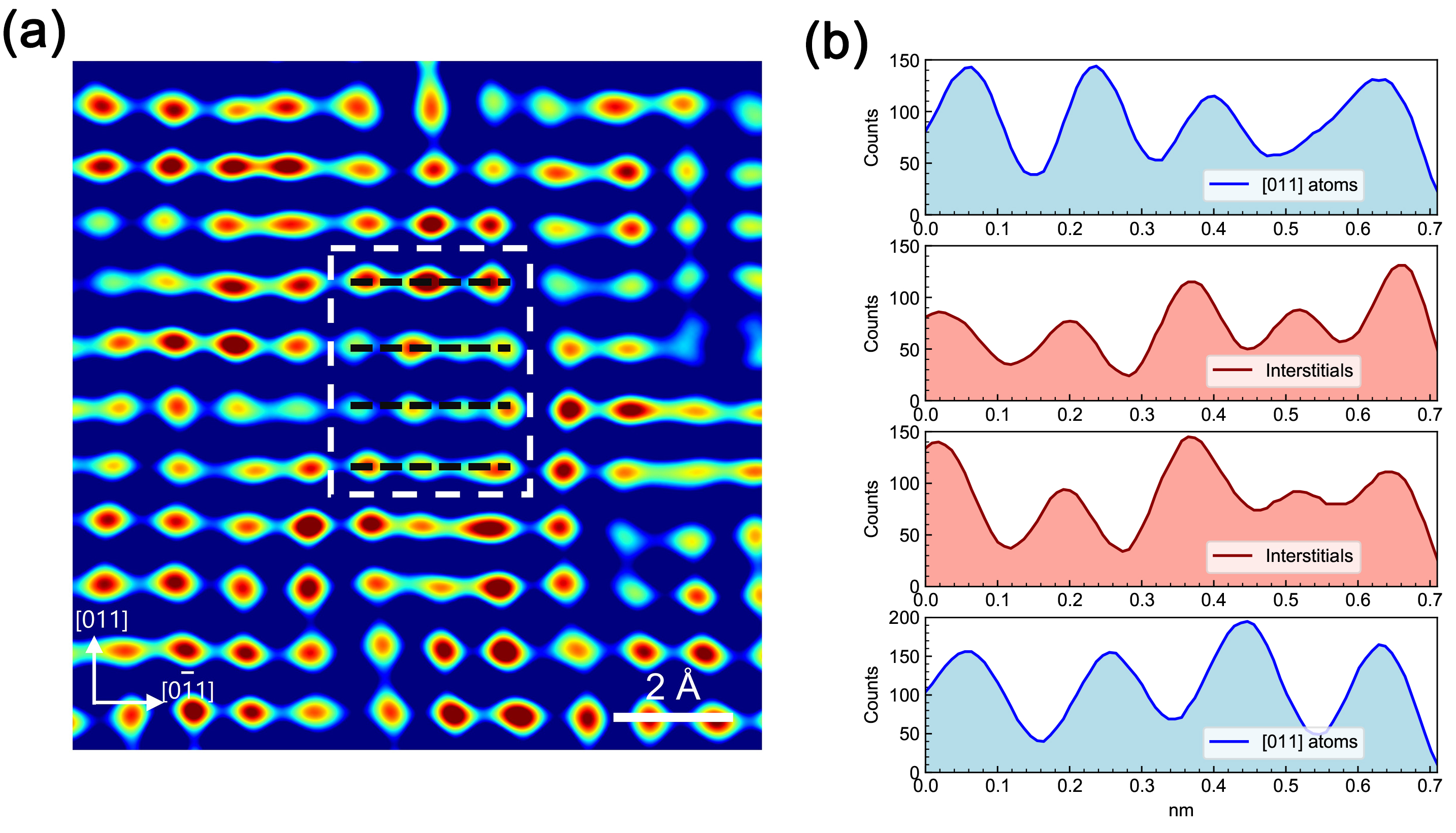}
\refstepcounter{appendixfigure}
\caption{\textbf{a,} Color map of FFT filtered HAADF image of dumbbell interstitials in diamond [$\overline{1}00$] view; 
         \textbf{b,} Line profile of interstitials in \{100\} a within the dashed box in \textbf{a} the blue represents the normal arrangement of diamond lattice in [$\overline{1}00$] view, and the red in the middle represents the projection of interstitial atoms. 
         }
\end{figure}

\begin{figure}[H]
\label{figS2}
\centering
\includegraphics[width=\linewidth]{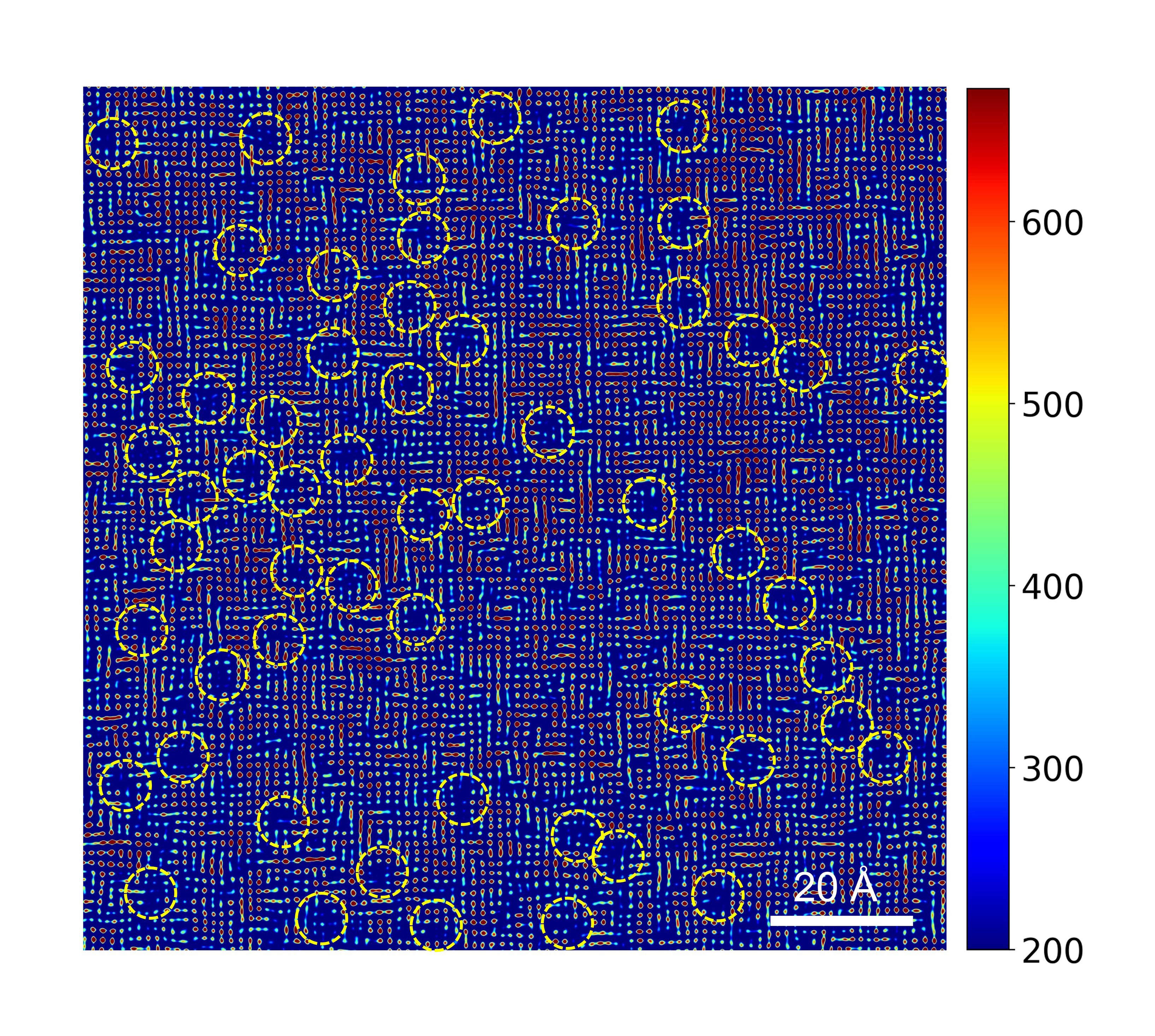}
\refstepcounter{appendixfigure}
\caption{Vacancy-type point defects in the HAADF image after Fourier filtering, indicated by reduced image intensity in atomic columns and marked with yellow circles..
         }
\end{figure}

\begin{figure}[H]
\label{figS3}
\centering
\includegraphics[width=\linewidth]{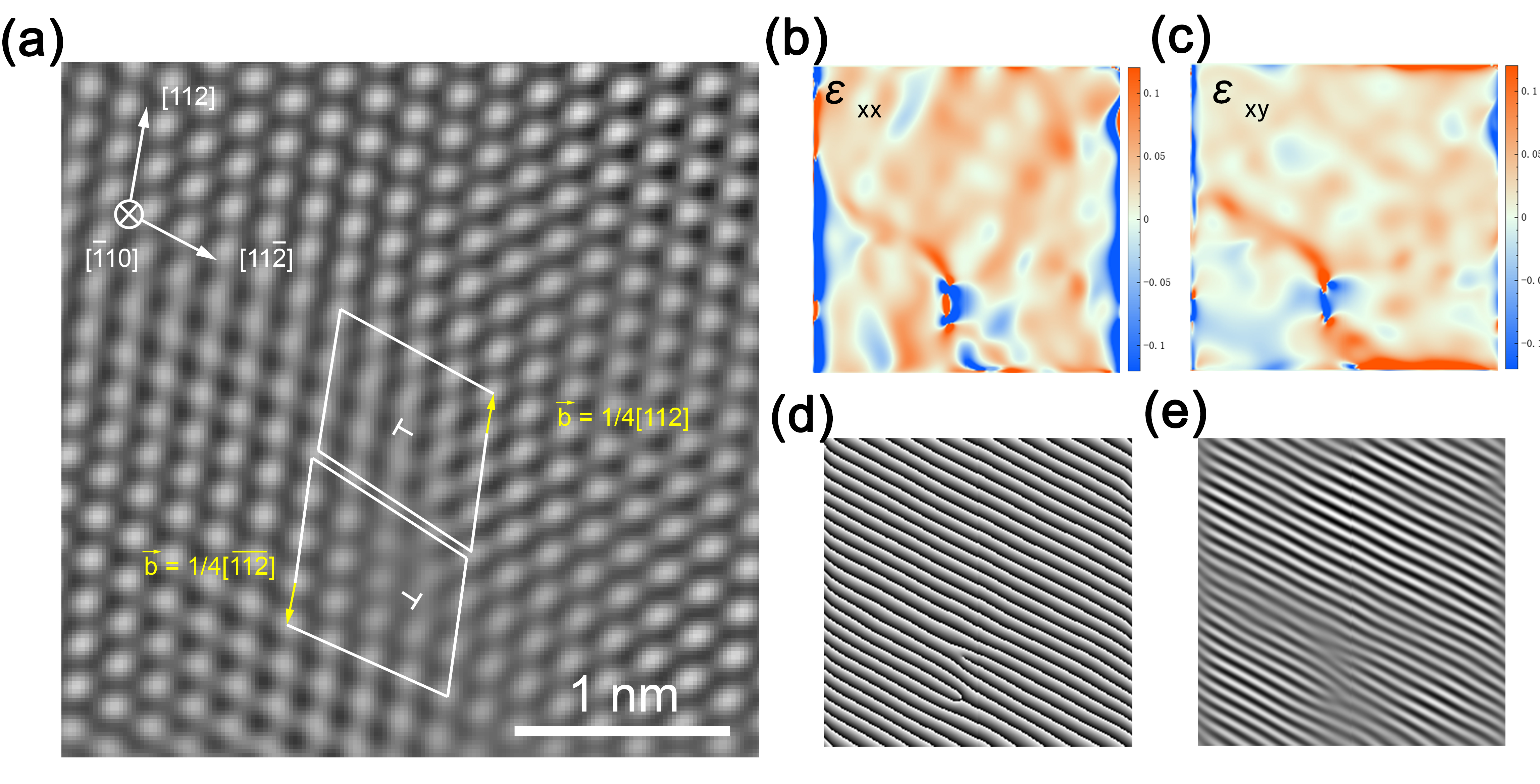}
\refstepcounter{appendixfigure}
\caption{\textbf{a,} Dipole (pair) dislocations with burgers vector of $\bm{b}=\frac{1}{4}[112]$ and $\bm{b}=\frac{1}{4}[\overline{112}]$ in $(\overline{1}10)$ diamond lattice plane; 
         \textbf{b} and \textbf{c,} Strain field of $\varepsilon_{xx}$ and $\varepsilon_{xy}$ of \textbf{a} based on GPA, several nanometer of positive and negative stresses are distributed in neutron radiation induced dislocations; 
         \textbf{d,} Raw phase of $(111)$ and $(\overline{111})$ lattice diffraction points from \textbf{a}, two opposite dislocation site are shown to separate by two $\{ 111\}$ lattice planes; 
         \textbf{e,} Bragg phase of $(111)$ and $(\overline{111})$ diamond lattice plane.}
\end{figure}

\begin{figure}[H]
\label{figS4}
\centering
\includegraphics[width=\linewidth]{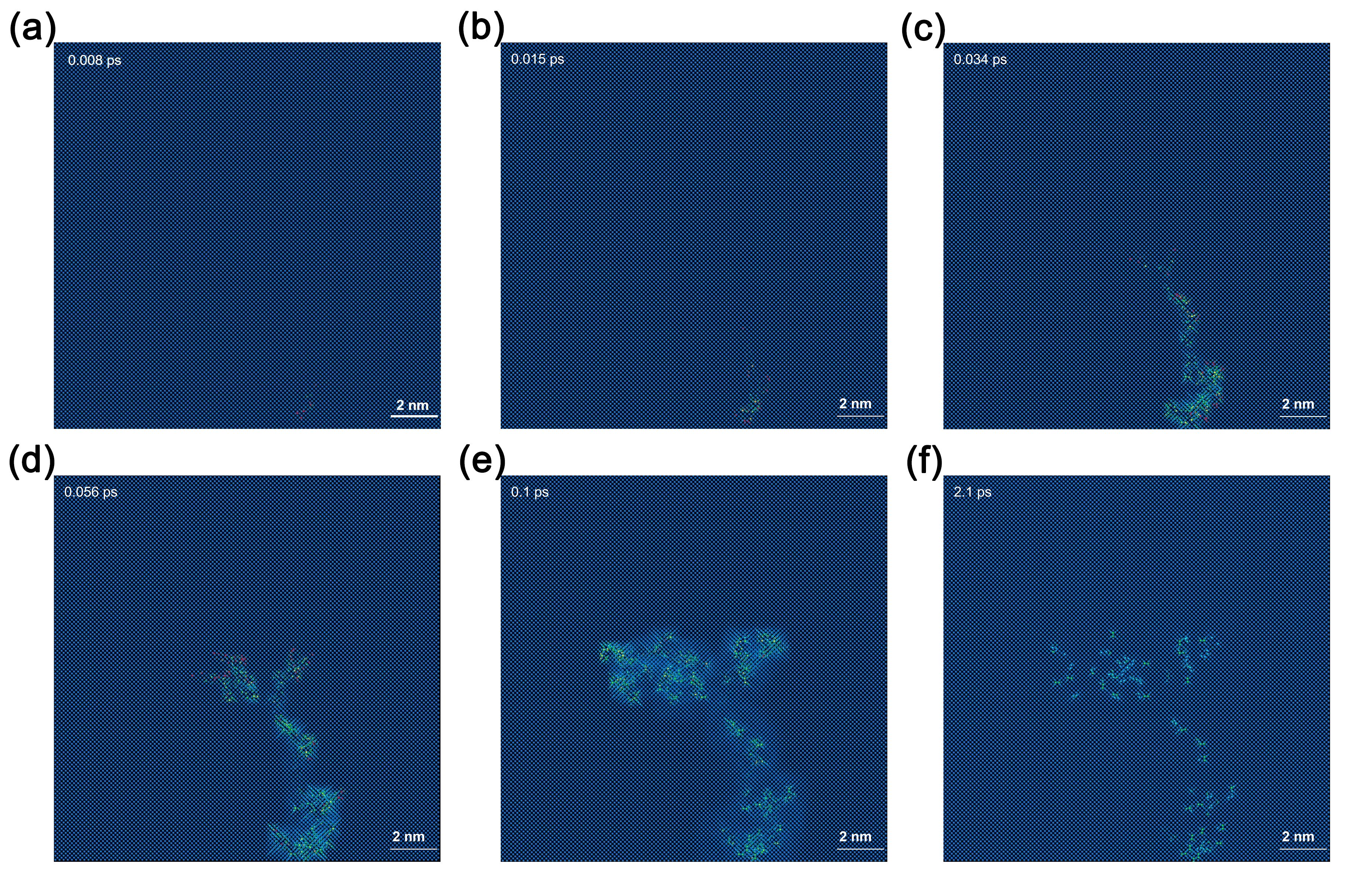}
\refstepcounter{appendixfigure}
\caption{Temporal evolution of the diamond lattice viewed from an orthogonal perspective following the incidence of a 7keV PKA. The snapshots depict the lattice at various time intervals: 0.008ps, 0.015ps, 0.034ps, 0.056ps, 0.1ps, and 2.1ps.}
\end{figure}

\begin{figure}[H]
\label{figS5}
\centering
\includegraphics[width=\linewidth]{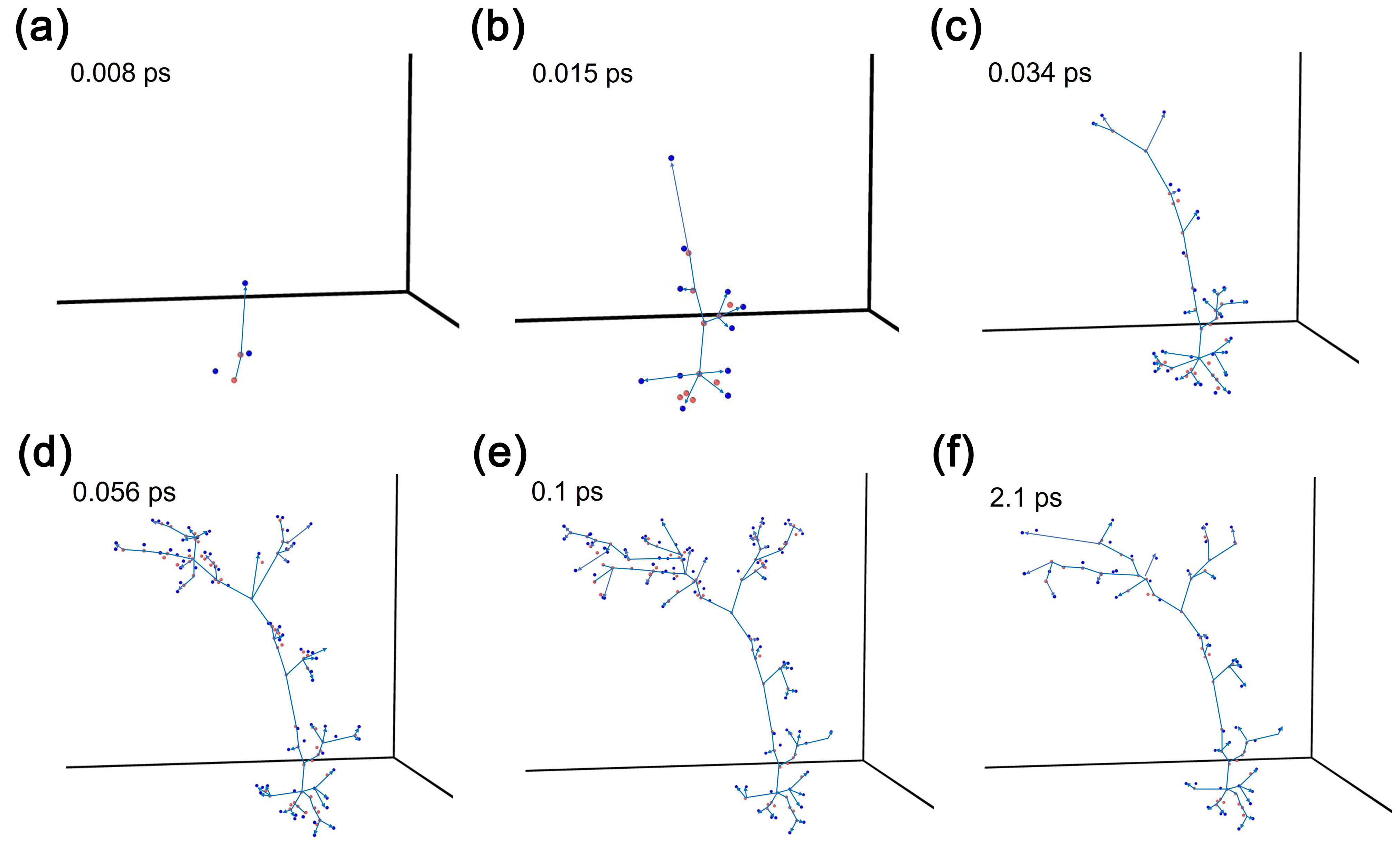}
\refstepcounter{appendixfigure}
\caption{The evolution of Frenkel defects in diamond over time upon irradiation by a 7keV PKA, where the interstitial defects are denoted by blue and vacancies by red. The blue lines depict the trajectory of the energetic recoil atoms, illustrating branches indicative of defect cluster formation during PKA migration, while interstitial atoms exhibit considerable ejection distances.}
\end{figure}

\end{document}